# Advisor career stage and PhD advisee outcomes

Xi Hong[1], Jialin Liu[1], Chaoqun Ni[1,*]
[1] Information School, College of Computing & Artificial Intelligence,
University of Wisconsin-Madison, Madison, WI, USA
[*]Corresponding author: chaoqun.ni@wisc.edu

**Abstract:** PhD advisors are central to doctoral training, but their influence may vary across career stages. Early-, mid-, and late-career advisors may differ in research activity, mentoring capacity, professional networks and access to resources. However, little is known about how PhD advisor career stage is associated with PhD student development outcomes. Drawing on multiple large-scale datasets comprising 250,838 advisor-advisee pairs from 312 U.S. PhD-granting institutions, we examine the relationship between advisor career stage and PhD advisee outcomes in knowledge production, collaboration networks and academic career placement. We find that early-career PhD advisors are associated with advisees' higher research productivity and citation performance, more opportunities to engage in direct and intensive research collaboration, and greater likelihood of securing a faculty position. Mid- and late-career faculty, by contrast, appear to have advantages in providing network capital which students can inherit after graduation, training PhD advisees to produce disruptive research, and supporting them in securing faculty positions at top institutions. This study contributes to a more comprehensive understanding of the reproduction of scientific talent by revealing the role of advisor career stage in shaping this process. These findings have implications for doctoral applicants' decision-making and for institutional policymaking on PhD training, faculty support and faculty evaluation.

## Introduction

Doctoral education plays a pivotal role in advancing science and innovation through training future scholars, with PhD advisors at the center of this training system in supporting student development and success. While factors such as student background and institutional environment also matter (Sverdlik et al., 2018), prior studies have widely recognized the role of advisors in shaping multiple dimensions of PhD advisee professional development and achievements, including research productivity and impact, collaboration formation, and career trajectories (Baruffaldi et al., 2016; Ma et al., 2020; Paglis et al., 2006; Rose & Shekhar, 2023). Such impact is often realized through advising efforts like advisory time input, quality and accountability, as well as their scholarly capital, such as research experience, resources and collaboration networks (García-Suaza et al., 2020; Heinisch & Buenstorf, 2018; C. E. Hilmer & Hilmer, 2007; Wang et al., 2021; Young-Jones et al., 2013). Moreover, beyond the direct efforts of mentoring, resources and professional opportunities, advisors can influence doctoral training outcomes through role modeling: By observing their advisors as visible examples of academic professionals, doctoral students develop professional norms, research and career aspirations, and approaches to publishing, collaboration and networking (Kram, 1983). Notably, PhD advisors at different career stages can differ substantially in the above features, potentially shaping PhD student development in different ways.

Research has documented the different characteristics of advisors at different career stages. A scholar's career stage generally refers to a scholar's position along the academic life course, commonly measured by years since their PhD or first publication, or by academic-rank-based milestones such as tenure status (APA, 2022; Robinson-Garcia et al., 2020; Tripodi et al., 2025). Existing research shows that early-career advisors are often under stronger publication incentives to survive the tenure evaluation (Tripodi et al., 2025), tend to devote more time and attention to individual mentoring and support, and are more directly involved in students' projects (Narendorf et al., 2016; Niles et al., 2020; Xing et al., 2025). Such commitment to mentoring is believed to be strongly associated with doctoral student success (Brill et al., 2014; Gottlieb, 1961; Green & Bauer, 1995; Slaughter & Zickar, 2006; Teschendorf & Nemshick, 2001). As advisors progress into mid- and late-career stages, they generally accumulate more research resources, broader professional networks, and stronger collaborative ties (M. J. Hilmer & Hilmer, 2009; Wang et al., 2021). From the perspective of doctoral socialization, these factors can advance advisees' socialization process in their professional field by providing more opportunities for research visibility, building collaborative connections, and facilitating integration into the community (Gardner & Barnes, 2007). Besides, senior scholars appear more inclined to place greater emphasis on innovative and breakthrough-oriented research agendas (Tripodi et al., 2025), which can help cultivate advisees' research taste. However, senior advisors may have less capacity for individualized mentoring, as they tend to lead larger research groups and are more likely to shoulder heavier administrative and leadership responsibilities (Malmgren et al., 2010; Rockwell, 2009; Xing et al., 2025). Overall, early-career advisors may provide closer supervision and stronger publication-oriented incentives, whereas mid- and late-career advisors may provide better resources, cultivation of research taste and networks. Therefore, the association between advisor career stage and PhD advisee development can be ambiguous and likely varies across dimensions. This brings up a key question: how is PhD advisor career stage associated with PhD student development outcomes, mainly reflected by different aspects of achievements, such as publication productivity, research impact, collaborations and career pathways?

Not enough research has discussed the relationship between advisor career stage and PhD advisee development outcomes, and existing studies are often restricted by limited sample sizes and narrow sets of outcomes (Corsini et al., 2022; Paglis et al., 2006). Moreover, existing research has yet to reach a consensus on this relationship, with contradictory findings supporting either the "Rising star" hypothesis, which supports the advantages brought by early-career advisors and the "Established star" hypothesis, which advocates for the advantages brought by established advisors (C. E. Hilmer & Hilmer, 2007; M. J. Hilmer & Hilmer, 2009; Malmgren et al., 2010).

Integrating multiple large-scale datasets, this study tries to contribute to this topic and provide implications for scientific workforce training and policymaking. Specifically, we explore the relationship between advisor career stage and PhD advisee development from three aspects, including knowledge production, academic career placement, and collaboration network formation based on 250,838 PhD advisor-advisee pairs from 312 U.S. Ph.D.-granting universities. Our sample includes PhD advisees who earned their PhD degrees between 2006 and 2019, spanning five broad fields: Engineering, Mathematics and Computing, Medicine and Health, Natural Sciences and Social Sciences.

## Results

**1. Early-career advisors are associated with advisee advantages in productivity and citations, while mid- and late-career advisors boost disruptive research for advisees who have strong disruptive-research potential.**

Our results indicate that PhD graduates supervised by advisors at different career stages exhibit different strengths in publication metrics during their PhD training period and within five years after their graduation. Advisees supervised by early-career advisors exhibit higher publication productivity and achieve more citations than those supervised by mid- and late-career advisors, while advisees mentored by late-career advisors show a higher level of research disruption in both periods (**Fig. 1A**). Specifically, compared with advisees mentored by early-career advisors, those mentored by mid- and late-career advisors exhibit differences of -0.038 (95% CI: [-0.043, -0.033]) and -0.060 (95% CI: [-0.066, -0.055]) in field- and year-normalized productivity, and -0.145 (95% CI: [-0.187, -0.103]) and -0.252 (95% CI: [-0.305, -0.200]) in field- and year-normalized citations during doctoral training, respectively, with these gaps persisting after graduation. By contrast, compared with advisees mentored by early-career advisors, those mentored by late-career advisors exhibit a 0.006 higher field- and year-normalized research disruption score (95% CI: [0.003, 0.010]) during doctoral training, and this difference is maintained after graduation. To verify whether these differences are mainly driven by explicit advisor publication performance, we further control for advisors' performance in productivity, citations and disruption in each model, respectively. The results show that the differences in citations remain significant but are smaller, the differences in disruption become insignificant, whereas the differences in productivity remain significant and even larger. Although the results for citation impact and research disruption support the role of advisors' observable research performance in accounting for the association between advisor career stage and student publication outcomes, the remaining significant differences in productivity and citations suggest that observable research performance cannot fully explain these associations. This finding points to the potential importance of other less observable mentoring processes, such as mentoring input and role-modeling effects.

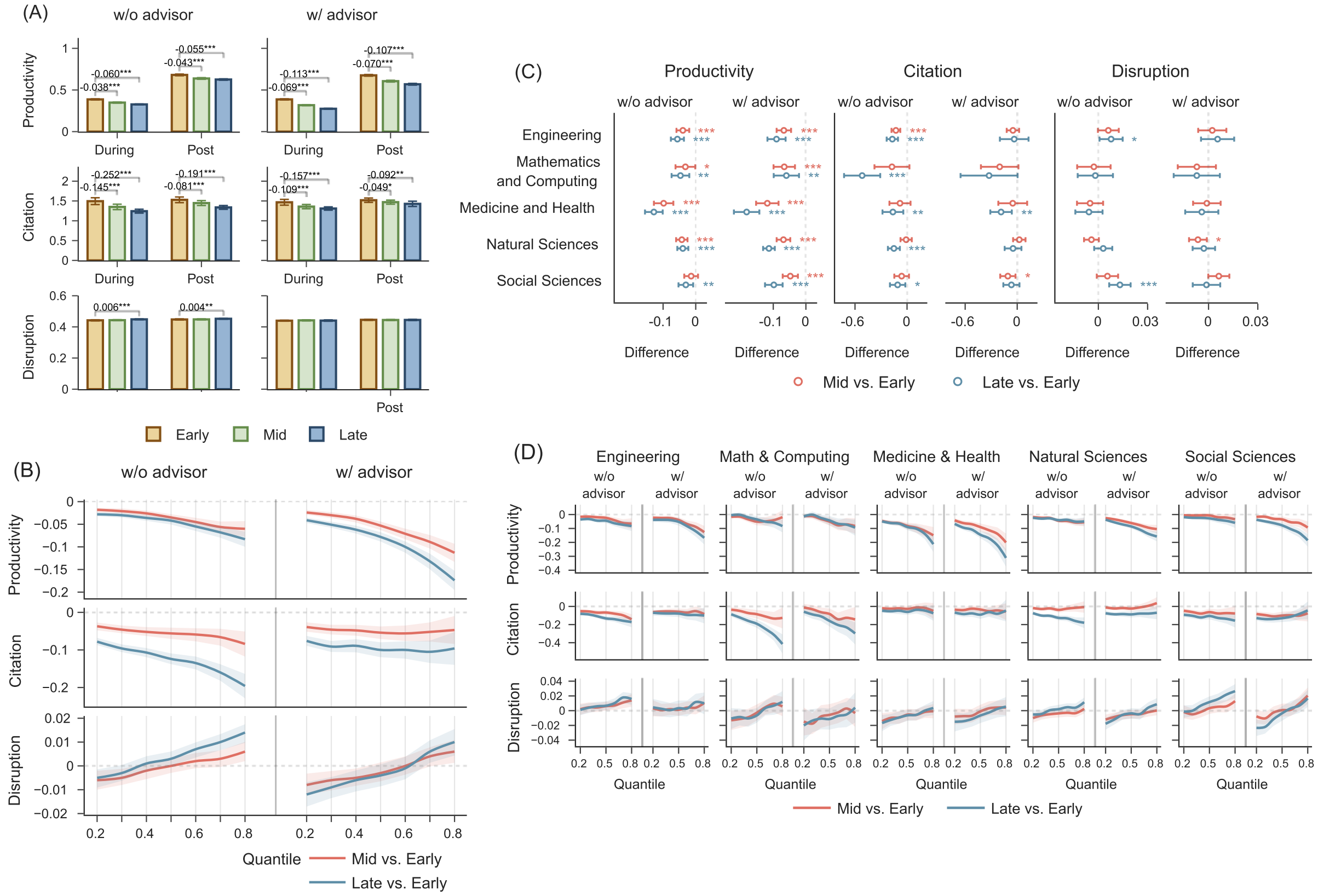

**Figure 1:** Publication performance differences between advisees mentored by mid-/late-career advisors and those mentored by early-career advisors in field- and year-normalized productivity, citations and disruption. (A) Differences between compared groups, during doctoral training ("During") and within five years after PhD graduation ("Post"). The bars are the adjusted means using post-estimation margins, while the numeric labels are differences estimated by regressions. (B) Differences between compared groups within 5 years after their PhD graduation, by quantile of the full sample (C) Differences between compared groups within 5 years after their PhD graduation, by discipline. (D) Differences between compared groups within 5 years after their PhD graduation, by discipline and quantile. Notably, for the 'post' period, we only consider advisees who have at least one publication record within five years after graduation to reduce the bias caused by advisees who have left academia or research settings. For quantile regressions, moving from the 0.2 to the 0.8 quantile represents examining students located progressively higher in the distribution of each metric. Thus, higher quantiles correspond to students with relatively stronger performance on that metric. "w/o advisor" refers to models that do not control for advisor publication metrics, while "w/ advisor" denotes models that control for advisor publication metrics. The plot centers represent the difference coefficients, while the endpoints or shaded areas represent the 95% confidence intervals. Significance level: * $P < 0.05$, ** $P < 0.01$, *** $P < 0.001$.

We further use quantile regressions to examine whether the heterogeneity of student characteristics will lead to different patterns of the relationship between advisor career stage and student publication outcome after their graduation **(Fig. 1B)**. Moving from lower to higher quantiles, the models capture the relationship across the distribution of student performance, from lower-performing to higher-performing students in each metric. The results show that the differences in productivity and citations between advisees mentored by early-career advisors and those mentored by mid- and late-career advisors are more pronounced at higher quantiles than at lower quantiles. Specifically, compared with advisees mentored by early-career advisors, those mentored by mid- and late-career advisors exhibit differences of -0.018 (95% CI: [-0.023, -0.014]) and -0.028 (95% CI: [-0.033, -0.023]) in field- and year-normalized productivity, and -0.037 (95% CI: [-0.048, -

0.027]) and -0.078 (95% CI: [-0.087, -0.068]) in field- and year-normalized citations at the 0.2 quantile, respectively. At the 0.8 quantile, these differences widen to -0.060 (95% CI: [-0.078, -0.043]) and -0.083 (95% CI: [-0.100, -0.066]) for productivity, and -0.084 (95% CI: [-0.118, -0.051]) and -0.196 (95% CI: [-0.229, -0.164]) for citations, respectively. This pattern suggests that as student capabilities or propensities related to productivity and citations increase, the advantage associated with being supervised by an early-career advisor rather than a mid- or late-career advisor can become larger. Moreover, PhD advisees supervised by mid- and late-career advisors perform better in research disruption than those supervised by early-career advisors, but only at the higher quantile groups. At lower quantiles, however, PhD students mentored by early-career advisors show better performance in research disruption. The above patterns imply the importance of distinguishing the heterogeneity of student features when exploring such relationships: the association between advisor career stage and the innovation level of students' research can be different given differentiated student characteristics.

**Figures 1C and 1D** exhibit differences between compared groups within 5 years after their PhD graduation. The results show consistency across disciplines that students mentored by early-career advisors perform better in productivity than those mentored by mid- or late-career advisors, which is still robust after advisor publication performance is controlled for. Besides, PhD advisees supervised by mid- and late-career advisors, particularly late-career advisors, tend to achieve higher research disruption scores in Engineering, Natural Sciences and Social Sciences. However, this advantage is significant at higher quantiles and can be largely explained by advisors' own levels of research disruption. Moreover, while all disciplines share a similar pattern as the total sample that students supervised by early-career advisors show an advantage in citations, the enlarged advantage at higher student quantiles is not obvious in Medicine and Health, Natural Sciences and Social Sciences.

**2. Early-career advisors are associated with a higher likelihood of advisee placement at U.S. institutions, whereas mid- and late-career advisors are associated with placement at higher-ranked institutions.**

Our findings show that advisor career stage is associated with different advisee placement outcomes across overall faculty placement and placement at higher-ranked institutions. PhD advisees mentored by early-career advisors are more likely to secure a faculty position at U.S. institutions than those mentored by mid- or late-career advisors within five years after PhD graduation **(Fig 2A)**. Overall, advisees supervised by early-career advisors have 16.2% (95% CI: [12.3%, 20.0%]) and 27.5% (95% CI: [24.1%, 30.9%]) higher odds of obtaining a faculty position at U.S. institutions than those advised by mid- and late-career advisors, respectively. Specifically, compared with advisees mentored by late-career advisors, the difference associated with early-career advisors is observed in Engineering (29.9%, 95% CI: [23.0%, 36.1%]), Mathematics and Computing (15.9%, 95% CI: [5.3%, 25.3%]), Medicine and Health (41.1%, 95% CI: [33.1%, 48.2%]), Natural Sciences (27.8%, 95% CI: [20.8%, 34.2%]) and Social Sciences (24.7%). Compared with graduates advised by mid-career advisors, the corresponding differences are 17.1% (95% CI: [7.9%, 25.3%]) in Engineering, 30.8% (95% CI: [20.4%, 39.9%]) in Medicine and Health, 22.9% (95% CI: [14.5%, 30.5%]) in Natural Sciences, and 10.4% (95% CI: [2.4%, 17.8%]) in the Social Sciences. We further control for students' publication metrics, including productivity and citations during their doctoral training, to better examine differences in placement beyond those explained by the observable PhD research output. It indicates that the likelihood differences become smaller but remain significant in most groups, except that there is no significant difference between advisees mentored by early- and late-career advisors in Mathematics and Computing, and no significant difference between advisees mentored by early- and mid-career advisors in the Social Sciences. These findings suggest that advisor career stage is associated with advisee faculty placement, although part of this association is explained by students' publication performance during doctoral training.

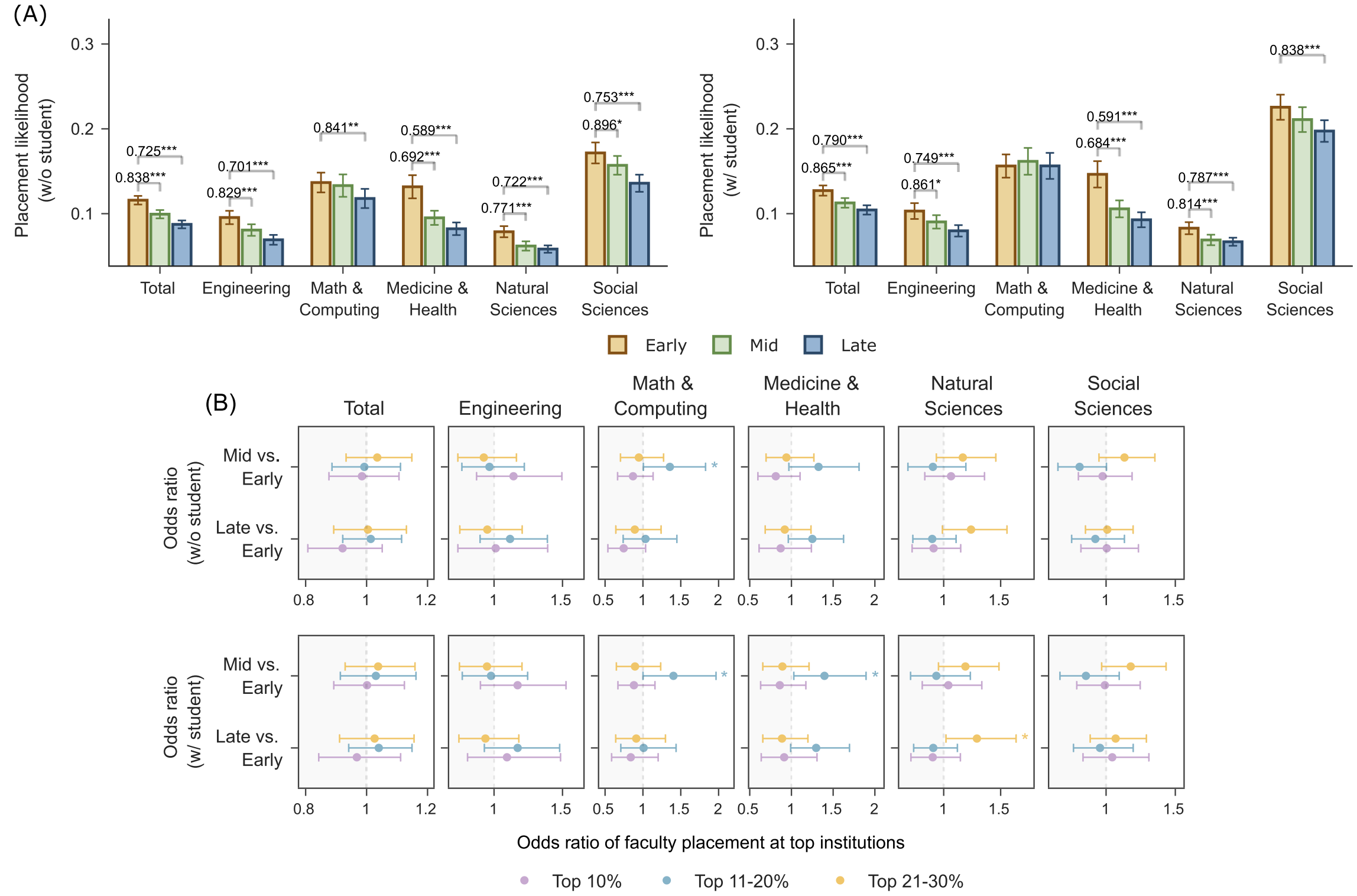

**Figure 2:** Differences in the likelihood of securing a faculty position at U.S. institutions within five years after PhD graduation, comparing advisees mentored by mid-/late-career advisors with those mentored by early-career advisors. **(A)** Differences in the overall likelihood of securing a faculty position at U.S. institutions between compared groups. The bars are the adjusted means estimated from the regression models using post-estimation margins, while the numeric labels are the odds ratios for comparisons between groups. **(B)** Differences in the likelihood of securing a faculty position at top 10%, top 11-20%, and top 21-30% U.S. institutions between compared groups. "w/o student" refers to models that do not control for student publication metrics, while "w/ student" denotes models that control for student publication metrics. The plot centers represent the odds ratio coefficients, while the endpoints represent the 95% confidence intervals. Significance level: * $P < 0.05$, ** $P < 0.01$, *** $P < 0.001$.

However, among graduates who secured faculty positions, we observe a contrasting pattern for advisee placement at higher-ranked institutions, with advantages associated with mid- and late-career advisors **(Fig. 2B)**. Without controlling for doctoral-period research output including productivity and citations, advisees mentored by mid-career advisors in Mathematics and Computing show higher odds of securing faculty positions at U.S. institutions ranked in the top 11-20% than those mentored by early-career advisors (OR = 1.356, 95% CI: [1.006, 1.828]). After controlling for doctoral-period research output, advisees mentored by mid-career advisors exhibit higher odds of placement at U.S. institutions ranked in the top 11-20% in Mathematics and Computing (OR = 1.403, 95% CI: [1.001, 1.966]) and Medicine and Health (OR = 1.396, 95% CI: [1.029, 1.894]). In addition, advisees mentored by late-career advisors in Natural Sciences present higher odds of placement at U.S. institutions ranked in the top 21-30% (OR = 1.292, 95% CI: [1.020, 1.635]). This denotes that given similar research productivity and citations during the doctoral period, advisees mentored by mid- and late-career advisors are more likely to obtain a faculty position at top U.S. institutions in STEM fields.

**3. Advisees mentored by early-career advisors have more coauthors per paper, which is positively associated with their likelihood of securing a faculty position at U.S. institutions, while advisees mentored by late-career advisors are more likely to independently collaborate with their advisors' collaborators, especially senior collaborators, after PhD graduation.**

The full sample shows that PhD advisees mentored by early-career advisors tend to have more coauthors per paper than those mentored by mid- and late-career advisors during their PhD training period (**Fig. 3A**). It suggests that compared with advisees mentored by early-career advisors, those mentored by mid- and late-career advisors have fewer coauthors per paper, with a difference of -0.017 (95% CI: [-0.031, -0.003]) and -0.052 (95% CI: [-0.067, -0.036]), respectively. We further test how this pattern of coauthor intensity may relate to PhD advisee career placement and find that having more coauthors per paper during the PhD training period is associated with a greater likelihood of securing a faculty position within five years after graduation (OR = 1.110, 95% CI: [1.083, 1.137]), which is consistent across different disciplines (**Fig. 3B**).

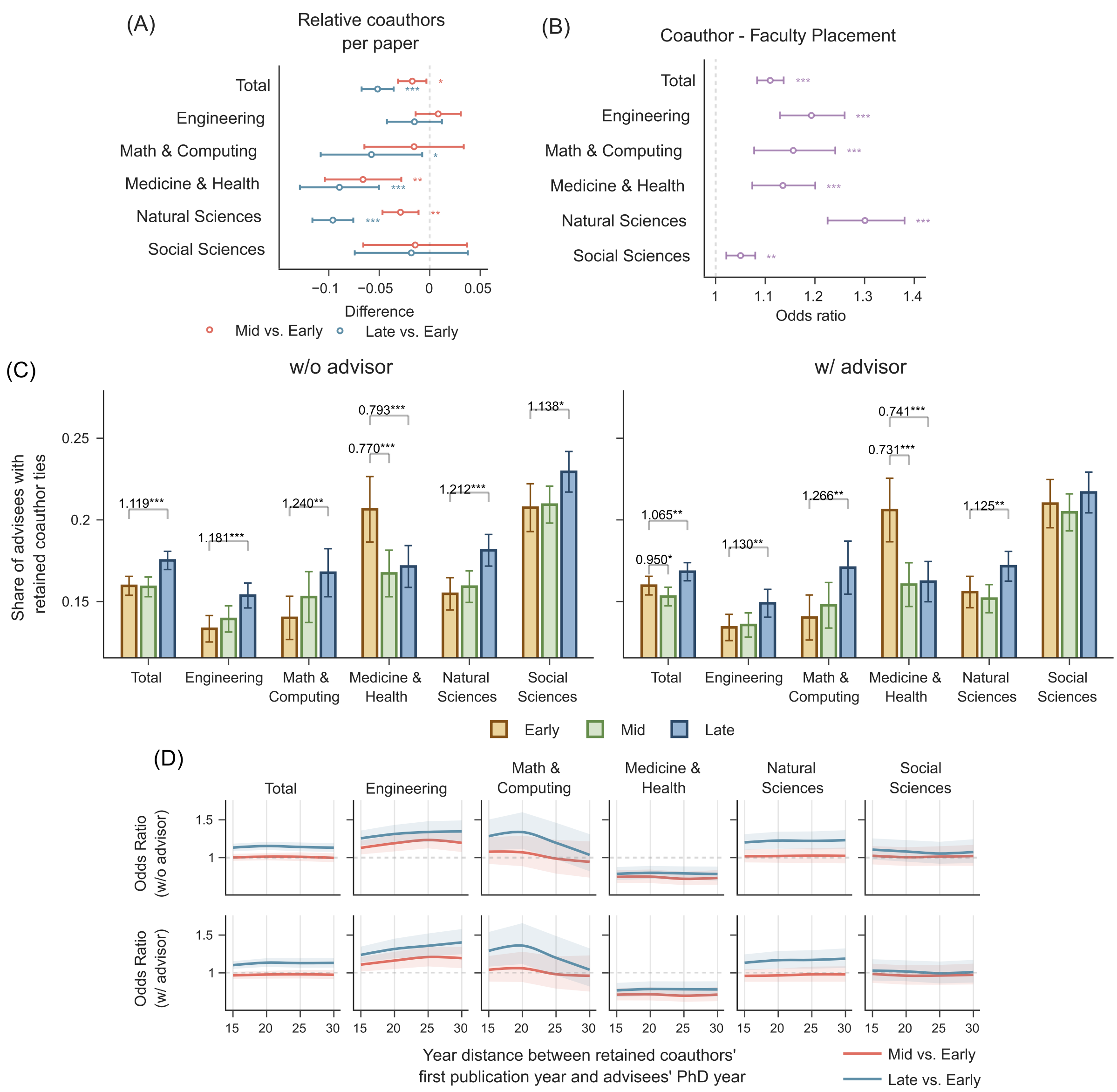

**Figure 3:** PhD advisee collaboration patterns by advisor career stage. **(A)** Differences in field- and year-normalized coauthors per paper during PhD training, comparing advisees mentored by mid-/late-career advisors with those mentored by early-career advisors. **(B)** Association between field- and year-normalized coauthors per paper and the likelihood of securing a faculty position at U.S. institutions within five years after PhD graduation. **(C)** Differences in the share of advisees inheriting advisors' collaborator ties within five years after PhD graduation, comparing advisees mentored by mid-/late-career advisors with those mentored by early-career advisors. The bars are the adjusted means estimated from the regression models using post-estimation margins, while the numeric labels are the odds ratios for comparisons between groups. **(D)** Differences in the share of advisees inheriting advisors' senior collaborator ties within five years after PhD graduation, comparing advisees mentored by mid-/late-career advisors with those mentored by early-career advisors. The plot centers represent the difference or odds ratio coefficients, while the endpoints or shaded areas represent the 95% confidence intervals. * $P < 0.05$, ** $P < 0.01$, *** $P < 0.001$.

On the other hand, PhD advisees mentored by late-career advisors show a higher probability of inheriting their advisors' collaborator ties after graduation. This is reflected in their greater likelihood of independently collaborating with at least one coauthor from their advisors' collaborator networks after graduation without their advisors' direct involvement. Specifically, students mentored by late-career advisors are more likely to inherit such ties than those mentored by early-career advisors in Engineering (1.181, 95% CI: [1.100, 1.269]), Mathematics and Computing (1.240, 95% CI: [1.078, 1.426]), Natural Sciences (1.212, 95% CI: [1.120, 1.312]) and Social Sciences (1.138, 95% CI: [1.023, 1.266]), respectively. The pattern remains consistent after controlling for advisor collaboration metrics, except in the Social Sciences. Besides, there is no significant difference between PhD graduates mentored by mid- and early-career advisors in Engineering, Mathematics and Computing, Natural Sciences, and Social Sciences (**Fig. 3C**).

We further measure the seniority of the collaborators based on the year window between the given collaborator's first publication and the given PhD advisee's graduation year, as well as the retention of these senior collaborators (**Fig. 3D**). The results indicate a consistent pattern across Engineering, Mathematics and Computing and Natural Sciences: after graduating from their PhD programs, advisees mentored by late-career advisors have a higher tendency to collaborate independently with the senior scholars in their advisors' collaborator ties, without advisors' direct involvement. Besides, advisees supervised by mid-career advisors show a higher senior collaborator retention rate than those supervised by early-career advisors in Engineering.

Notably, in Medicine and Health, the pattern is reversed: advisees mentored by early-career advisors are more likely to retain their advisors' collaborator ties after graduation. One possible explanation is the distinctive collaborative structure of this field: research such as clinical trials and related data-intensive projects often requires data sharing, coordinated teamwork and sustained collaboration (Bennett & Gadlin, 2012; NIH, 2015). This is supported by our data that Medicine and Health has the highest average raw number of coauthors per paper for both advisors and advisees across all disciplines (**Supplementary Table 1**). Given advisees mentored by early-career advisors have more coauthors per paper in this field as shown above, they can be more likely to inherit at least one collaboration tie of their advisors after PhD graduation.

## Discussions

This study highlights the role of advisor career stage in producing the next generation of the scientific workforce from the perspectives of knowledge production, academic career placement and network formation. It shows that compared to PhD advisees supervised by mid- and late-career advisors, those supervised by early-career advisors exhibit higher publication productivity and greater citation impact in their research. Conversely, those mentored by mid-and late-career advisors show higher research disruption, but this association is evident primarily among advisees who demonstrate strong propensities for conducting disruptive research. Besides, PhD students supervised by early-career advisors have more coauthors listed on their papers during doctoral training, which is associated with a higher probability of obtaining a U.S. faculty position, while advisees supervised by late-career advisors are more likely to retain their advisors' collaborator ties, especially senior collaborators, after PhD graduation. Moreover, advisees supervised by early-career advisors are more likely to land a faculty job at U.S. institutions, but students mentored by mid- and late-career advisors show a higher probability of securing an academic job at top U.S. institutions. These findings have implications for doctoral applicants' decision-making, as well as institutional policymaking related to PhD training and faculty support and evaluation.

For doctoral program applicants, our findings suggest that there is no universal pattern of 'better' PhD outcomes tied to advisor career stage. Both advantages and tradeoffs can happen when choosing advisors at different career stages: while early-career advisors appear to provide collaboration and productivity intensive training and a higher likelihood of academic career placement, senior advisors can transmit research taste, network capital, and prestige in academic career placement. Moreover, advisor-advisee matching in PhD applications is a complex process that requires careful consideration of students' most pressing needs, as well as reflection on their own characteristics and goals. For example, our study finds that for students with different propensities to produce disruptive research, the advantages associated with advisors at different career stages may vary substantially. Therefore, doctoral applicants need to make application decisions based not only on advisors' seniority, but also on the type of training, support, and developmental opportunities that best fit their own demands, aspirations and strengths.

To better support PhD advising, institutions need to provide more targeted support to advisors and seek approaches to integrate the strengths of advisors at different career stages. As this study shows, early-career PhD advisors are playing an important role in training future faculty with higher productivity and citation impact. At the same time, they often face competing demands from research, teaching, service, tenure evaluation and PhD advising. Therefore, institutions should provide targeted resources, such as mentoring training and startup funding for doctoral student training, while reducing administrative and service burdens where possible. Moreover, tenure and other faculty evaluations should give clearer credit to PhD advising to recognize and encourage faculty contributions to the training of future scholars. Besides, as our findings indicate that advisors at different career stages may possess different strengths for student development outcomes, institutions may consider redesigning the PhD advising system by encouraging and incorporating more joint mentoring, such as a co-advising model between junior and senior faculty to maximize the strengths and impact of advisors at different career stages.

## Conclusions and limitations

Integrating multiple large-scale scholarly datasets and 250,838 PhD advisor-advisee pairs at 312 U.S. institutions, our study shows that in the U.S. context, early-career PhD advisors are associated with advisees' higher research productivity and citation performance, more opportunities to engage in direct and intensive research collaboration, and greater likelihood of securing a faculty position. Mid- and late-career faculty, by contrast, appear to have advantages in providing network capital which students can inherit after graduation, training PhD advisees to produce disruptive research, and supporting them in securing faculty positions at top institutions. The findings of this study contribute to a more comprehensive understanding of the reproduction of scientific talent by revealing the role of advisor career stage in shaping this process.

This study has limitations. First, the paper-based publication profiles may not fully capture research activities in fields where other outputs like books and patents also play an important role. We have tried to mitigate this by excluding disciplines like Humanities. Second, despite our use of CEM to improve comparability across advisor career-stage groups, the analyses may still be subject to biases such as the unobserved time-invariant differences between advisors. As a next step, we plan to extend the advisor-advisee data prior to 2006 to implement advisor fixed-effects models as a robustness check. Third, the publication-based indicators adopted in this study capture important dimensions, but they may not encompass other aspects like broader industrial and societal impact. Moreover, although we examine collaboration patterns through collaborator intensity and ties retention, our analysis lacks a network perspective to better understand how advisors at different career stages shape advisees' integration into scientific networks. Finally, our analysis of career placement concentrates on U.S. academia, which can not fully capture the career outcomes of PhD advisees. As doctoral graduates increasingly pursue careers globally and in non-academic sectors, future research should examine a broader range of career trajectories to provide a more comprehensive view of doctoral training outcomes.

**Research data and methods**

**Data.** This study encompasses four data sources: the ProQuest doctoral-dissertation dataset, containing the information on 930,354 PhD dissertations completed at U.S. institutions from 2006-2024 (D1); Academic Analytics Research Center (AARC) faculty rosters, covering 310,303 tenured and tenure-track faculty at 393 U.S. Ph.D.-granting institutions from 2011 to 2020 (D2); OpenAlex, a comprehensive bibliometric database of scientific papers, authors and institutions, maintained by OurResearch (D3); and SciSciNet, a large-scale open data lake covering over 134 million research publications (Lin et al., 2023) (D4).

**Matching advisor-advisee pairs.** Based on D1 and D2, we generate PhD advisor-advisee pairs by matching the PhD primary advisor names and institutions in D1 with faculty names and institutions in D2. We restrict the sample to selected disciplines, including Engineering, Mathematics and Computing, Medicine and Health, Natural Sciences and Social Sciences (**see Supplementary Note 1 for discipline classification details**). This selection is based on two considerations: (1) research outcomes in disciplines like Humanities are often not paper-based, whereas the data we use to construct scholars' publication profiles are largely paper-based; and (2) in fields such as Education, research-oriented doctoral programs also include a substantial number of professional degrees, particularly the Ed.D., which fall outside the scope of this study. Moreover, we focus on PhD advisees who graduated before 2020 to allow a reasonable time window for the accumulation of publication records. Based on the above criteria, we finally identify 355,583 advisor-advisee pairs, covering 85,472 unique PhD advisors.

**Matching publication profiles.** We link the advisor-advisee pairs to D3 to build comprehensive publication records for each advisor and advisee. Specifically, for advisors, we link D2 and D3 via publications' Digital Object Identifiers (DOIs) to obtain the advisors' OpenAlex author IDs. For advisees, we adopt a multi-step matching approach. First, we use the dissertation year, title, advisee name and institution to match records between D1 and D3. Given that these features might not be consistently recorded in the two datasets, we apply the following matching criteria in sequence: (1) all four variables; (2) all variables except institution; (3) all variables except the advisee's name; and (4) all variables except the dissertation title. Moreover, we attempt to match the advisee's name to the author lists of their advisor's publication records as an additional matching strategy. Only records matched by at least one approach are included in the analytic sample. When a record matched multiple approaches, we applied the strictest matching criterion (**see Supplementary Tab. 2 for details**). After these steps, 270,151 unique advisor-advisee pairs are successfully matched with publication records in OpenAlex. To assess potential name-disambiguation errors arising from our matching procedures, we conduct a manual validation of the matched records. Five researchers, each with more than five years of research experience, participate in the verification process, with each reviewer independently evaluating 500 unique records. The overall matching accuracy is 95.7%, indicating a high level of matching quality and reliability. We further exclude cases in which a single OpenAlex ID is linked to multiple ProQuest records or person IDs, and cases where the advisor's PhD completion year is later than the advisee's entry year plus one. After applying these filters, the sample consists of 254,955 advisor–advisee pairs. The full match rates across different graduation years and disciplines are provided in **Supplementary Table 3**. Moreover, to reduce the potential under-splitting of OpenAlex IDs, we further restrict the sample by applying one-sided trimming, which excludes observations in the top 1% of the distribution. We also re-estimate all analyses using both untrimmed data and one-sided trimmed data that excludes observations in the top 2.5% of the distribution as a robustness check (**Supplementary Fig. 1-6)**. After the above steps, the final sample includes 250,838 unique advisor-advisee pairs with advisee PhD graduation years from 2006 to 2019, including 71,458 unique advisors and 250,838 advisees from 312 U.S. PhD-granting institutions. Based on the sample, we retrieve a total of 14,042,339 publication records from OpenAlex for the focal authors, including information on complete authorship lists, citation counts, and the primary disciplines of the publications. We further identify each author's first and most recent publication years in OpenAlex and obtain paper-level disruption scores by matching the publications to SciSciNet.

**Key Variables.** ***Advisor career stage.*** Given an advisor-advisee pair, we classify each advisor's career stage as early, mid, or late based on the number of years since their PhD completion during the advisee's doctoral training period. If, for the majority of the advisee's training years, the advisor is within 10 years of PhD graduation, they are categorized as early-career; those within 20 years as mid-career; and those beyond 20 years as late-career (NIH, 2024; APA, 2022; AGU, 2026). Advisees' PhD degree years are obtained from D1, while advisors' PhD degree years are drawn from D2. Because PhD entry years are not directly observable, we estimate them using the median time to degree reported in the NSF Survey of Earned Doctorates (NSF, 2023). Specifically, each advisee's entry year was estimated as five years before their PhD degree year. ***Publication metrics.*** We assess the publication performance of PhD advisees by three metrics: productivity, citations and disruption. (1) ***Productivity*** is measured by the relative total publications, defined as an individual's total number of publications normalized by the average publication count of scholars in the same field and year. (2) ***Citations*** are measured by the year-and field-normalized C5 metric in D3, referring to the number of citations a publication received in the five years following its publication time. (3) ***Disruption*****,** which indicates "the extent to which a paper disrupts or develops the existing literature", is measured by the citation networks (see Lin et al. (2023) for details) in D4. Notably, citations and disruption are measured as the average across an individual's publications. The time window for advisees' publication metrics includes two periods: during their PhD training period (the 'during' period) and within five years after their PhD graduation (the 'post' period). For the 'post' period, we only consider advisees who have at least one publication record within five years after graduation to reduce the bias caused by individuals who have left academia or research settings. Moreover, to reduce the publication overlap between advisor and advisee pairs when calculating the adviser's publication metrics for a given advisor-advisee pair, we define the advisor's publication window as the period from five years before the advisor's PhD graduation year to the year before the advisee's entry year. In particular, starting the window five years before PhD graduation provides a reasonable span for capturing advisors' publication records, especially if they are at an early-career stage. ***Faculty Placement.*** We assess faculty placement in U.S. academia among advisees who graduated between 2006 and 2015 based on whether they appeared in D2 as faculty within five years after PhD graduation (**See Supplementary Tab. 4 for details**). We further explore their placement in the institutions ranked top 10%, 11-20% and 21-30%, based on the SpringRank algorithm (De Bacco et al., 2018) (**Supplementary Note 2**). ***Collaborations*****.** We measure collaboration patterns by two approaches: 1. Relative collaborators per paper, defined as an individual's average number of coauthors per paper normalized by the field and year baseline. It denotes the pattern of collaboration intensity in each research project. 2. Collaborator ties inheritance, measured by whether an advisee continues collaborating after graduation, without the advisor's involvement, with coauthors shared by the advisor and advisee during the advisee's PhD training. Because most advisees do not retain such collaborations after graduation, we measure it by a dummy variable, with 0 denoting no continuous collaboration and 1 denoting continuous collaboration with at least one common coauthor. Moreover, we calculate the seniority of the collaborators based on whether the year window between the given collaborator's first publication and the given PhD advisee's graduation year exceeds 15, 20, 25 and 30 years, and calculate the inheritance of these senior collaborators, respectively.

**Statistical Analysis.** We employ several statistical analysis techniques, including Coarsened Exact Matching (CEM), quantile regressions, linear regressions and logistic regressions to examine the relationship between advisor career stage and student publication performance, collaborator ties inheritance and job placement. Specifically, CEM is adopted in all models to balance observed covariates across treatment and control groups such as student entry year, student pre-entry publication record, institution rank, and advisor publication and collaboration metrics (see **Supplementary Note 3 and Tab. 5 for CEM details and balance check**).

We include several covariates to strengthen the regression design, including advisee PhD entry year, advisees' publication records before they entered their PhD program, discipline and institution rank.

For Q1, we use linear regressions and quantile regressions to explore the relationship between advisor career stage and PhD advisee publication performance.

$$Y_i = \alpha + \beta_M \, Mid_i + \beta_L \, Late_i + X_i'\gamma + \varepsilon_i$$

where $Y_i$ represents the advisee publication-performance outcome; $Mid_i$ and $Late_i$ indicate mid-career and late-career advisors, with early-career advisors as the reference group; $X_i$ represents covariates; and $\varepsilon_i$ is the error term.

$$Q_\tau(Y_i;\, Mid_i, Late_i, X_i) = \alpha_\tau + \beta_{M\tau} \, Mid_i + \beta_{L\tau} \, Late_i + X_i'\gamma_\tau$$

where $Q_\tau(Y_i;\, Mid_i, Late_i, X_i)$ represents the conditional $\tau$-th quantile of advisee publication performance, and the $\tau$-specific coefficients describe associations at different points of the outcome distribution.

For Q2, linear regressions are adopted to predict  the association between advisor career stage and PhD advisees' collaboration intensity per paper, whereas logistic regressions are applied for predicting the relationship between advisor career stage and PhD advisees' inheritance of collaborator ties.

$$C_i = \alpha + \beta_M \, Mid_i + \beta_L \, Late_i + X_i'\gamma + \varepsilon_i$$

where $C_i$ represents the advisee collaboration-intensity outcome per project; $Mid_i$, $Late_i$, and $X_i$ are defined as above; and $\varepsilon_i$ is the error term.

$$logit\, Pr(H_i = 1 \mid Mid_i, Late_i, X_i) = \alpha + \beta_M \, Mid_i + \beta_L \, Late_i + X_i'\gamma$$

where $H_i$ equals 1 if the advisee inherits at least one advisor collaborator tie after PhD graduation, and 0 otherwise; the model estimates the log-odds of collaborator-tie inheritance.

For Q3, logistic regressions are adopted to explore the probability of landing a faculty position at U.S. institutions.

$$logit\, Pr(F_i = 1 \mid Mid_i, Late_i, X_i) = \alpha + \beta_M \, Mid_i + \beta_L \, Late_i + X_i'\gamma$$

where $F_i$ equals 1 if the advisee obtains a faculty position at U.S. institutions, and 0 otherwise; the model estimates the log-odds of U.S. faculty placement.

Other robustness or sensitivity checks include: (1) junior and senior career stages classified by the predicted tenure year of advisors using machine learning models as a robustness check for the three-class advisor career stage classification (**Supplementary Note 4 and Fig. 7-9**); (2) "hit papers" as a sensitivity check for the impact-oriented publication productivity, defined as the number of an individual's publications that fall within the top 5% of the citation distribution in a given field and year (**Supplementary Note 5 and Fig. 10**); (3) Beta regressions as a robustness check for the collaborator ties inheritance: for advisees who inherited at least one collaborator tie of their advisors after PhD graduation, we further examine the proportion of the collaborators they retained (a continuous fractional variable between 0 and 1) and compare the differences across groups using Beta regressions (**Supplementary Fig. 11**).

**Data, Materials, and Software Availability.** Some study data are available. Dataset D1, licensed data from ProQuest (https://www.proquest.com), is accessible by visiting https://tdmstudio.proquest.com. Dataset D2, a licensed research dataset from the Academic Analytics Research Center (AARC), is accessible by contacting AARC directly at https://aarcresearch.com. Dataset D3, a comprehensive bibliometric database of scientific papers, authors and institutions, is accessible by visiting https://openalex.org. Dataset D4, SciSciNet, is accessible by visiting https://huggingface.co/datasets/Northwestern-CSSI/sciscinet-v2/tree/main.

**Supplementary file**

**Supplementary Note 1.** Discipline classification

We assign faculty to academic fields based on the AARC's original classification of faculty departments, which includes 187 granular taxonomies. From these taxonomies, we categorize each faculty member into one of seven broad fields: Engineering, Mathematics and Computing, Medicine and Health, Natural Sciences, Social Sciences, Education and Humanities.

**Supplementary Note 2.** University ranking by the SpringRank algorithm

We apply the SpringRank algorithm (De Bacco et al., 2018) to faculty hiring networks within each AARC subfield to infer university rankings. Using faculty Ph.D. and employment affiliations, SpringRank assigns a continuous prestige score to each university-subfield pair, which we convert to rank percentiles.

**Supplementary Note 3.** Coarsened exact matching (CEM)

We implement three rounds of CEM to improve comparability among students mentored by early-, mid-, and late-career advisors. In each round, every matching stratum is required to contain at least one observation from each career-stage group, and post-matching weights are constructed within strata so that the weighted group totals are anchored to those of the early-career group. The first round matches on basic pretreatment characteristics, including the student's pre-PhD publications, field, PhD entry year, and institution rank. The second round additionally matches on the advisor's publication metrics, whereas the third round additionally matches on the advisor's collaboration metrics.

For each CEM round, we assess covariate balance before and after matching using the sample with complete data on all variables included in that round as the pre-matching baseline and the weighted matched sample as the post-matching sample. **Supplementary Table 5** presents the results of the covariate balance checks for all three rounds. Because the comparison involves three career-stage groups rather than a binary treatment-control contrast, balance is summarized using the maximum absolute pairwise difference across the early-, mid-, and late-career groups. For continuous covariates, this statistic is reported as a standardized mean difference calculated using the standard deviation of the pre-matching sample. For binary and categorical covariates, we report the maximum absolute pairwise difference in proportions.

**Supplementary Note 4.** Advisor career stage based on the pre- and post-tenure period

We conduct a robustness check for the three-level advisor career stage classification using the pre-tenure and post-tenure classification predicted by machine learning models. To estimate the tenured year of faculty who have no clear record in our 2006-2020 AARC dataset, we train machine learning models to predict faculty tenure-period length, defined as the number of years from PhD degree year to the transition from assistant to associate professor. The predictors include faculty placement institution percentile, degree-granting institution percentile, gender, field, and degree year. Numeric variables are median-imputed and standardized, while categorical variables are imputed with the most frequent category and one-hot encoded. The training sample encompasses 62,845 faculty with clear observed transition years in the dataset. The data is split into 80% training and 20% test sets, and model tuning is conducted on the training set using 5-fold cross-validation with root mean squared error as the selection criterion. We compare mean-baseline, linear regression, regularized linear models, k-nearest neighbors, and random forest models. The best-performing model is a random forest with 800 trees, unrestricted tree depth, max features = 0.7, and minimum leaf size of 1, achieving a test RMSE of 1.82 years, MAE of 1.35 years, and $R^2$ of 0.856.

**Supplementary Note 5.** "Hit papers" as an impact-oriented research productivity measure

We adopt the "hit papers", an impact-oriented research productivity measure, as a sensitivity check for publication productivity, defined as the number of an individual's publications that fall within the top 5% of the citation distribution in a given field and year. Especially, we use raw counts of hit papers and apply Poisson regressions to predict the relationship between advisor career stage and advisee hit-paper performance.

**Supplementary Table 1.** Raw average number of coauthors per paper of advisors and advisees, by discipline

| Umbrella | Variable | Mean | SD | Min | 25% | 50% | 75% | Max |
|---|---|---|---|---|---|---|---|---|
| Engineering | Advisee coauthors per paper (during) | 1.90 | 1.83 | 0.00 | 0.60 | 1.40 | 2.75 | 15.83 |
| | Advisee coauthors per paper (post) | 1.96 | 2.04 | 0.00 | 0.20 | 1.40 | 3.03 | 14.98 |
| | Advisor coauthors per paper (Pre) | 1.83 | 1.05 | 0.00 | 1.12 | 1.62 | 2.34 | 12.66 |
| Mathematics and Computing | Advisee coauthors per paper (during) | 1.15 | 1.45 | 0.00 | 0.00 | 0.60 | 1.70 | 14.33 |
| | Advisee coauthors per paper (post) | 1.39 | 1.76 | 0.00 | 0.00 | 0.80 | 2.12 | 14.25 |
| | Advisor coauthors per paper (Pre) | 1.29 | 0.89 | 0.00 | 0.67 | 1.13 | 1.71 | 7.50 |
| Medicine and Health | Advisee coauthors per paper (during) | 2.40 | 2.29 | 0.00 | 0.60 | 1.80 | 3.60 | 16.80 |
| | Advisee coauthors per paper (post) | 2.94 | 2.66 | 0.00 | 0.60 | 2.40 | 4.68 | 16.20 |
| | Advisor coauthors per paper (Pre) | 2.38 | 1.27 | 0.00 | 1.48 | 2.29 | 3.13 | 10.54 |
| Natural Sciences | Advisee coauthors per paper (during) | 2.34 | 2.13 | 0.00 | 0.80 | 1.80 | 3.40 | 19.78 |
| | Advisee coauthors per paper (post) | 2.64 | 2.44 | 0.00 | 0.60 | 2.13 | 4.12 | 15.97 |
| | Advisor coauthors per paper (Pre) | 2.30 | 1.16 | 0.00 | 1.52 | 2.13 | 2.87 | 12.00 |
| Social Sciences | Advisee coauthors per paper (during) | 0.88 | 1.36 | 0.00 | 0.00 | 0.30 | 1.20 | 15.63 |
| | Advisee coauthors per paper (post) | 1.13 | 1.60 | 0.00 | 0.00 | 0.40 | 1.62 | 14.88 |
| | Advisor coauthors per paper (Pre) | 1.06 | 0.85 | 0.00 | 0.47 | 0.86 | 1.43 | 8.57 |

**Supplementary Table 2.** Approaches to matching advisee publication records in D3

| Step | Criteria | Percentage |
|---|---|---|
| Step 1 | Dissertation Title & Name & Institution & Year | 6.4% |
| Step 2 | Dissertation Title & Name & Year | 61.2% |
| Step 3 | Dissertation Title & Institution & Year | 0.2% |
| Step 4 | Dissertation Institution & Name & Year | 3.0% |
| Step 5 | Mentorship data | 29.1% |

**Supplementary Table 3.** Match rates of advisees in D3 by PhD graduation year and discipline

| Year | Engineering | Mathematics and Computing | Medicine and Health | Natural Sciences | Social Sciences |
|---|---|---|---|---|---|
| 2006 | 81.9% | 74.0% | 60.1% | 74.6% | 45.7% |
| 2007 | 84.5% | 73.6% | 66.8% | 77.5% | 49.7% |
| 2008 | 85.9% | 76.3% | 69.1% | 79.4% | 51.7% |
| 2009 | 85.3% | 79.1% | 72.3% | 81.8% | 59.5% |
| 2010 | 87.2% | 79.9% | 76.8% | 84.1% | 64.0% |
| 2011 | 87.4% | 80.7% | 76.4% | 84.1% | 64.5% |
| 2012 | 87.7% | 81.2% | 78.5% | 86.5% | 67.2% |
| 2013 | 89.0% | 82.1% | 80.8% | 85.6% | 69.8% |
| 2014 | 88.4% | 81.9% | 79.9% | 86.1% | 68.7% |
| 2015 | 88.9% | 81.8% | 80.5% | 86.8% | 70.5% |
| 2016 | 88.1% | 80.5% | 82.4% | 86.3% | 71.1% |
| 2017 | 85.4% | 79.1% | 75.3% | 83.7% | 64.6% |
| 2018 | 82.8% | 76.4% | 75.0% | 81.5% | 61.2% |
| 2019 | 71.5% | 60.5% | 68.0% | 72.1% | 49.0% |

**Supplementary Table 4.** U.S. faculty placement by discipline, advisor career stage and advisee graduation year

| Discipline | Graduation year | N (Early) | Placed (Early) | Rate (Early) | N (Mid) | Placed (Mid) | Rate (Mid) | N (Late) | Placed (Late) | Rate (Late) | All stages rate |
|---|---|---|---|---|---|---|---|---|---|---|---|
| Engineering | 2006 | 852 | 82 | 9.6% | 1,446 | 113 | 7.8% | 1,226 | 94 | 7.7% | 8.2% |
| | 2007 | 856 | 84 | 9.8% | 1,436 | 124 | 8.6% | 1,308 | 107 | 8.2% | 8.8% |
| | 2008 | 829 | 78 | 9.4% | 1,364 | 112 | 8.2% | 1,300 | 94 | 7.2% | 8.1% |
| | 2009 | 877 | 101 | 11.5% | 1,433 | 125 | 8.7% | 1,358 | 101 | 7.4% | 8.9% |
| | 2010 | 1,003 | 108 | 10.8% | 1,475 | 126 | 8.5% | 1,617 | 106 | 6.6% | 8.3% |
| | 2011 | 1,000 | 88 | 8.8% | 1,474 | 129 | 8.8% | 1,711 | 117 | 6.8% | 8.0% |
| | 2012 | 696 | 63 | 9.1% | 1,023 | 97 | 9.5% | 1,206 | 85 | 7.0% | 8.4% |
| | 2013 | 1,200 | 99 | 8.2% | 1,474 | 111 | 7.5% | 2,051 | 146 | 7.1% | 7.5% |
| | 2014 | 1,160 | 124 | 10.7% | 1,493 | 111 | 7.4% | 1,969 | 120 | 6.1% | 7.7% |
| | 2015 | 1,126 | 90 | 8.0% | 1,497 | 131 | 8.8% | 2,112 | 168 | 8.0% | 8.2% |
| Mathematics and Computing | 2006 | 335 | 51 | 15.2% | 474 | 67 | 14.1% | 441 | 57 | 12.9% | 14.0% |
| | 2007 | 313 | 39 | 12.5% | 475 | 64 | 13.5% | 459 | 54 | 11.8% | 12.6% |
| | 2008 | 347 | 53 | 15.3% | 461 | 59 | 12.8% | 451 | 53 | 11.8% | 13.1% |
| | 2009 | 405 | 58 | 14.3% | 556 | 80 | 14.4% | 558 | 71 | 12.7% | 13.8% |
| | 2010 | 399 | 59 | 14.8% | 567 | 74 | 13.1% | 657 | 81 | 12.3% | 13.2% |
| | 2011 | 433 | 49 | 11.3% | 565 | 77 | 13.6% | 645 | 69 | 10.7% | 11.9% |

| Discipline | Graduation year | N (Early) | Placed (Early) | Rate (Early) | N (Mid) | Placed (Mid) | Rate (Mid) | N (Late) | Placed (Late) | Rate (Late) | All stages rate |
|---|---|---|---|---|---|---|---|---|---|---|---|
| | 2012 | 355 | 46 | 13.0% | 391 | 56 | 14.3% | 585 | 71 | 12.1% | 13.0% |
| | 2013 | 506 | 67 | 13.2% | 660 | 81 | 12.3% | 772 | 91 | 11.8% | 12.3% |
| | 2014 | 454 | 49 | 10.8% | 587 | 76 | 12.9% | 704 | 82 | 11.6% | 11.9% |
| | 2015 | 463 | 77 | 16.6% | 626 | 72 | 11.5% | 742 | 69 | 9.3% | 11.9% |
| Medicine and Health | 2006 | 253 | 24 | 9.5% | 634 | 53 | 8.4% | 669 | 54 | 8.1% | 8.4% |
| | 2007 | 307 | 44 | 14.3% | 744 | 57 | 7.7% | 789 | 57 | 7.2% | 8.6% |
| | 2008 | 347 | 47 | 13.5% | 763 | 57 | 7.5% | 865 | 66 | 7.6% | 8.6% |
| | 2009 | 372 | 54 | 14.5% | 799 | 73 | 9.1% | 1,027 | 76 | 7.4% | 9.2% |
| | 2010 | 407 | 50 | 12.3% | 973 | 99 | 10.2% | 1,174 | 112 | 9.5% | 10.2% |
| | 2011 | 430 | 58 | 13.5% | 968 | 97 | 10.0% | 1,133 | 93 | 8.2% | 9.8% |
| | 2012 | 314 | 41 | 13.1% | 739 | 81 | 11.0% | 940 | 72 | 7.7% | 9.7% |
| | 2013 | 474 | 59 | 12.4% | 1,122 | 122 | 10.9% | 1,360 | 112 | 8.2% | 9.9% |
| | 2014 | 483 | 64 | 13.3% | 1,042 | 92 | 8.8% | 1,276 | 118 | 9.2% | 9.8% |
| | 2015 | 463 | 66 | 14.3% | 1,065 | 105 | 9.9% | 1,308 | 118 | 9.0% | 10.2% |
| Natural Sciences | 2006 | 716 | 48 | 6.7% | 1,745 | 132 | 7.6% | 2,015 | 134 | 6.7% | 7.0% |
| | 2007 | 775 | 66 | 8.5% | 1,705 | 88 | 5.2% | 2,149 | 138 | 6.4% | 6.3% |
| | 2008 | 765 | 71 | 9.3% | 1,733 | 97 | 5.6% | 2,108 | 154 | 7.3% | 7.0% |
| | 2009 | 912 | 73 | 8.0% | 1,980 | 113 | 5.7% | 2,495 | 169 | 6.8% | 6.6% |
| | 2010 | 962 | 58 | 6.0% | 2,077 | 143 | 6.9% | 2,767 | 135 | 4.9% | 5.8% |
| | 2011 | 1,068 | 89 | 8.3% | 1,991 | 116 | 5.8% | 2,755 | 164 | 6.0% | 6.3% |
| | 2012 | 704 | 57 | 8.1% | 1,509 | 103 | 6.8% | 2,175 | 145 | 6.7% | 7.0% |
| | 2013 | 1,052 | 84 | 8.0% | 2,234 | 156 | 7.0% | 3,152 | 171 | 5.4% | 6.4% |
| | 2014 | 1,013 | 87 | 8.6% | 2,289 | 147 | 6.4% | 3,018 | 177 | 5.9% | 6.5% |
| | 2015 | 1,040 | 76 | 7.3% | 2,299 | 123 | 5.4% | 3,136 | 182 | 5.8% | 5.9% |
| Social Sciences | 2006 | 445 | 65 | 14.6% | 817 | 146 | 17.9% | 1,003 | 144 | 14.4% | 15.7% |
| | 2007 | 474 | 82 | 17.3% | 837 | 132 | 15.8% | 1,132 | 182 | 16.1% | 16.2% |
| | 2008 | 515 | 85 | 16.5% | 921 | 159 | 17.3% | 1,206 | 205 | 17.0% | 17.0% |
| | 2009 | 670 | 119 | 17.8% | 1,094 | 184 | 16.8% | 1,588 | 234 | 14.7% | 16.0% |
| | 2010 | 731 | 123 | 16.8% | 1,343 | 198 | 14.7% | 1,810 | 233 | 12.9% | 14.3% |
| | 2011 | 761 | 119 | 15.6% | 1,251 | 198 | 15.8% | 1,901 | 250 | 13.2% | 14.5% |
| | 2012 | 668 | 142 | 21.3% | 1,152 | 204 | 17.7% | 1,682 | 284 | 16.9% | 18.0% |
| | 2013 | 927 | 155 | 16.7% | 1,472 | 255 | 17.3% | 2,275 | 332 | 14.6% | 15.9% |
| | 2014 | 854 | 157 | 18.4% | 1,503 | 242 | 16.1% | 2,152 | 312 | 14.5% | 15.8% |
| | 2015 | 862 | 137 | 15.9% | 1,544 | 237 | 15.3% | 2,236 | 319 | 14.3% | 14.9% |

## **Supplementary Table 5.** CEM balance check

| **Round 1: Basic covariates** | | | | | | | | | | |
|---|---|---|---|---|---|---|---|---|---|---|
| **Covariate** | **Checking approach** | **Early Before** | **Mid Before** | **Late Before** | **Max Imbalance Before** | **Early After** | **Mid After** | **Late After** | **Max Imbalance After** | **Reduction (%)** |
| Student entry year | SMD | 2008.08 | 2007.94 | 2008.17 | 0.06 | 2008.08 | 2008.08 | 2008.08 | 0.00 | 100.00 |
| Institution rank percentile | SMD | 73.59 | 75.12 | 77.32 | 0.17 | 73.59 | 73.62 | 73.80 | 0.01 | 94.53 |
| Student pre-PhD paper | PPD | 21.40 | 23.30 | 23.80 | 2.40 | 21.40 | 21.40 | 21.40 | 0.00 | 100.00 |
| Engineering | PPD | 29.40 | 23.20 | 21.90 | 7.50 | 29.40 | 29.40 | 29.40 | 0.00 | 100.00 |
| Mathematics and Computing | PPD | 11.60 | 9.00 | 8.00 | 3.60 | 11.60 | 11.60 | 11.60 | 0.00 | 100.00 |
| Medicine and Health | PPD | 11.80 | 14.80 | 14.00 | 3.00 | 11.80 | 11.80 | 11.80 | 0.00 | 100.00 |
| Natural Sciences | PPD | 26.60 | 32.60 | 34.00 | 7.40 | 26.60 | 26.60 | 26.60 | 0.00 | 100.00 |
| Social Sciences | PPD | 20.60 | 20.40 | 22.10 | 1.70 | 20.60 | 20.60 | 20.60 | 0.00 | 100.00 |
| **Round 2: Basic covariates and advisor publication metrics** | | | | | | | | | | |
| **Covariate** | **Checking approach** | **Early Before** | **Mid Before** | **Late Before** | **Max Imbalance Before** | **Early After** | **Mid After** | **Late After** | **Max Imbalance After** | **Reduction (%)** |
| Student entry year | SMD | 2008.15 | 2007.96 | 2008.18 | 0.06 | 2008.20 | 2008.20 | 2008.20 | 0.00 | 100.00 |
| Institution rank percentile | SMD | 74.18 | 75.38 | 77.48 | 0.16 | 74.02 | 74.05 | 74.34 | 0.02 | 90.29 |
| Advisor publications | SMD | 0.92 | 1.23 | 1.62 | 0.52 | 0.94 | 0.95 | 0.99 | 0.04 | 92.41 |
| Advisor citations | SMD | 1.98 | 1.70 | 1.45 | 0.36 | 1.94 | 1.76 | 1.61 | 0.22 | 38.21 |
| Advisor research disruption | SMD | 0.44 | 0.44 | 0.47 | 0.22 | 0.44 | 0.44 | 0.45 | 0.06 | 72.25 |
| Student pre-PhD paper | PPD | 21.90 | 23.40 | 23.90 | 2.00 | 19.00 | 19.00 | 19.00 | 0.00 | 100.00 |
| Engineering | PPD | 29.60 | 23.20 | 21.90 | 7.70 | 28.10 | 28.10 | 28.10 | 0.00 | 100.00 |
| Mathematics and Computing | PPD | 11.60 | 9.00 | 8.00 | 3.60 | 10.60 | 10.60 | 10.60 | 0.00 | 100.00 |
| Medicine and Health | PPD | 11.70 | 14.80 | 14.00 | 3.10 | 11.80 | 11.80 | 11.80 | 0.00 | 100.00 |
| Natural Sciences | PPD | 27.60 | 32.90 | 34.20 | 6.60 | 29.10 | 29.10 | 29.10 | 0.00 | 100.00 |
| Social Sciences | PPD | 19.50 | 20.10 | 21.90 | 2.40 | 20.40 | 20.40 | 20.40 | 0.00 | 100.00 |
| **Round 3: Basic covariates and advisor collaboration metrics** | | | | | | | | | | |
| **Covariate** | **Checking approach** | **Early Before** | **Mid Before** | **Late Before** | **Max Imbalance Before** | **Early After** | **Mid After** | **Late After** | **Max Imbalance After** | **Reduction (%)** |
| Student entry year | SMD | 2008.10 | 2007.95 | 2008.17 | 0.06 | 2008.10 | 2008.10 | 2008.10 | 0.00 | 100.00 |

| | | | | | | | | | |
|---|---|---|---|---|---|---|---|---|---|
| Institution rank percentile | SMD | 73.79 | 75.18 | 77.34 | 0.16 | 73.79 | 73.81 | 74.92 | 0.05 | 68.21 |
| Advisor coauthors per paper | SMD | 1.09 | 1.12 | 1.19 | 0.13 | 1.09 | 1.05 | 1.03 | 0.07 | 44.83 |
| Advisor collaborators per year | SMD | 3.92 | 4.85 | 5.90 | 0.29 | 3.92 | 4.06 | 4.21 | 0.04 | 85.72 |
| Student pre-PhD paper | PPD | 21.60 | 23.30 | 23.80 | 2.20 | 21.50 | 21.50 | 21.50 | 0.00 | 100.00 |
| Engineering | PPD | 29.60 | 23.30 | 21.90 | 7.60 | 29.60 | 29.60 | 29.60 | 0.00 | 100.00 |
| Mathematics and Computing | PPD | 11.60 | 9.00 | 8.00 | 3.60 | 11.60 | 11.60 | 11.60 | 0.00 | 100.00 |
| Medicine and Health | PPD | 11.70 | 14.80 | 14.00 | 3.00 | 11.70 | 11.70 | 11.70 | 0.00 | 100.00 |
| Natural Sciences | PPD | 27.10 | 32.70 | 34.00 | 6.90 | 27.10 | 27.10 | 27.10 | 0.00 | 100.00 |
| Social Sciences | PPD | 20.00 | 20.30 | 22.10 | 2.10 | 20.00 | 20.00 | 20.00 | 0.00 | 100.00 |

Notes: Continuous covariates report group means and max absolute pairwise standardized mean differences (SMD); binary and categorical covariates report percentages and max absolute pairwise percentage-point differences (PPD).

**Supplementary Figure 1-3: Robustness check using one-sided trimmed data that excludes observations in the top 2.5% of the distribution.**

**Supplementary Figure 1:** See Figure 1 for caption.

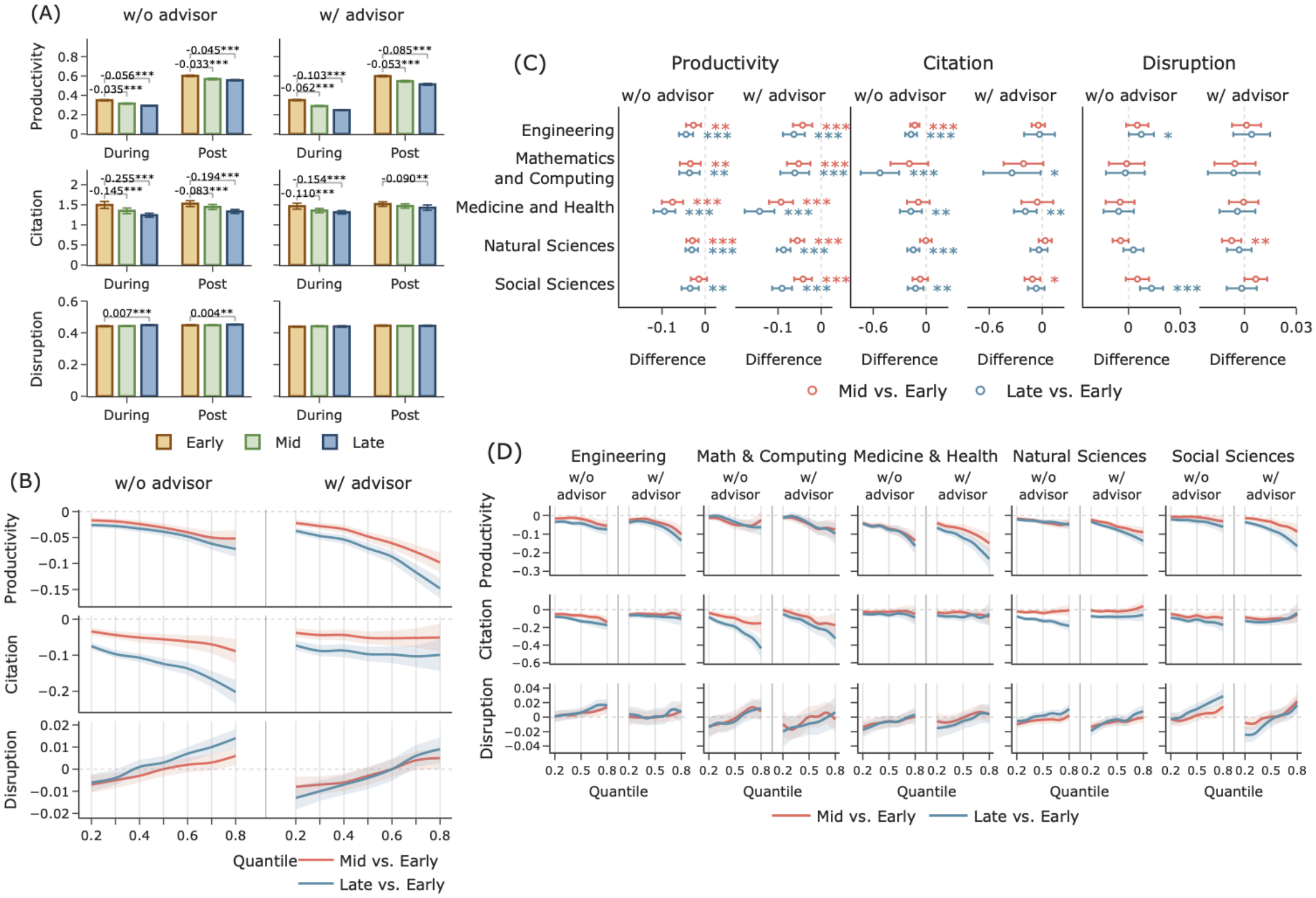

**Supplementary Figure 2:** See Figure 2 for caption.

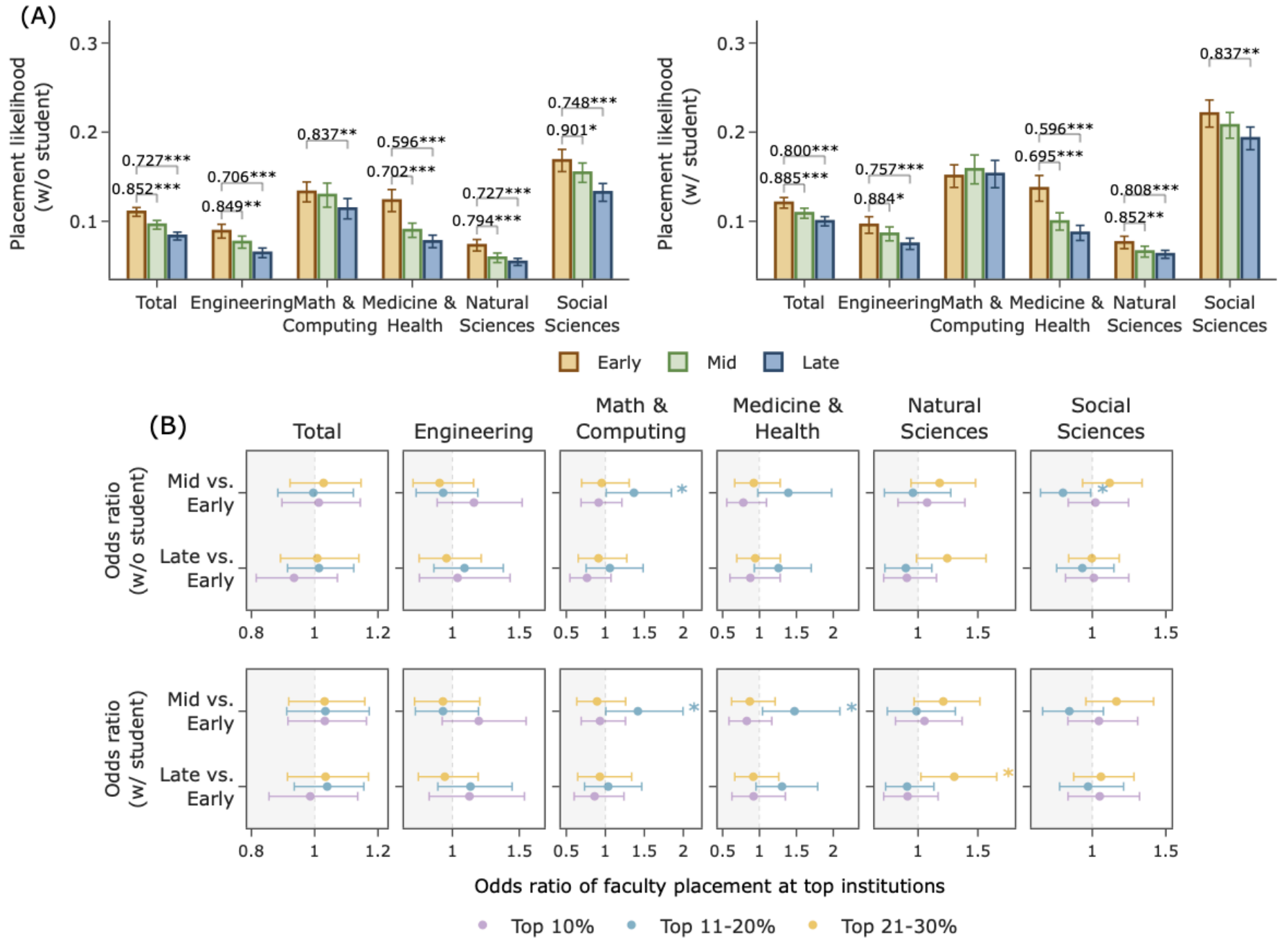

**Supplementary Figure 3:** See Figure 3 for caption.

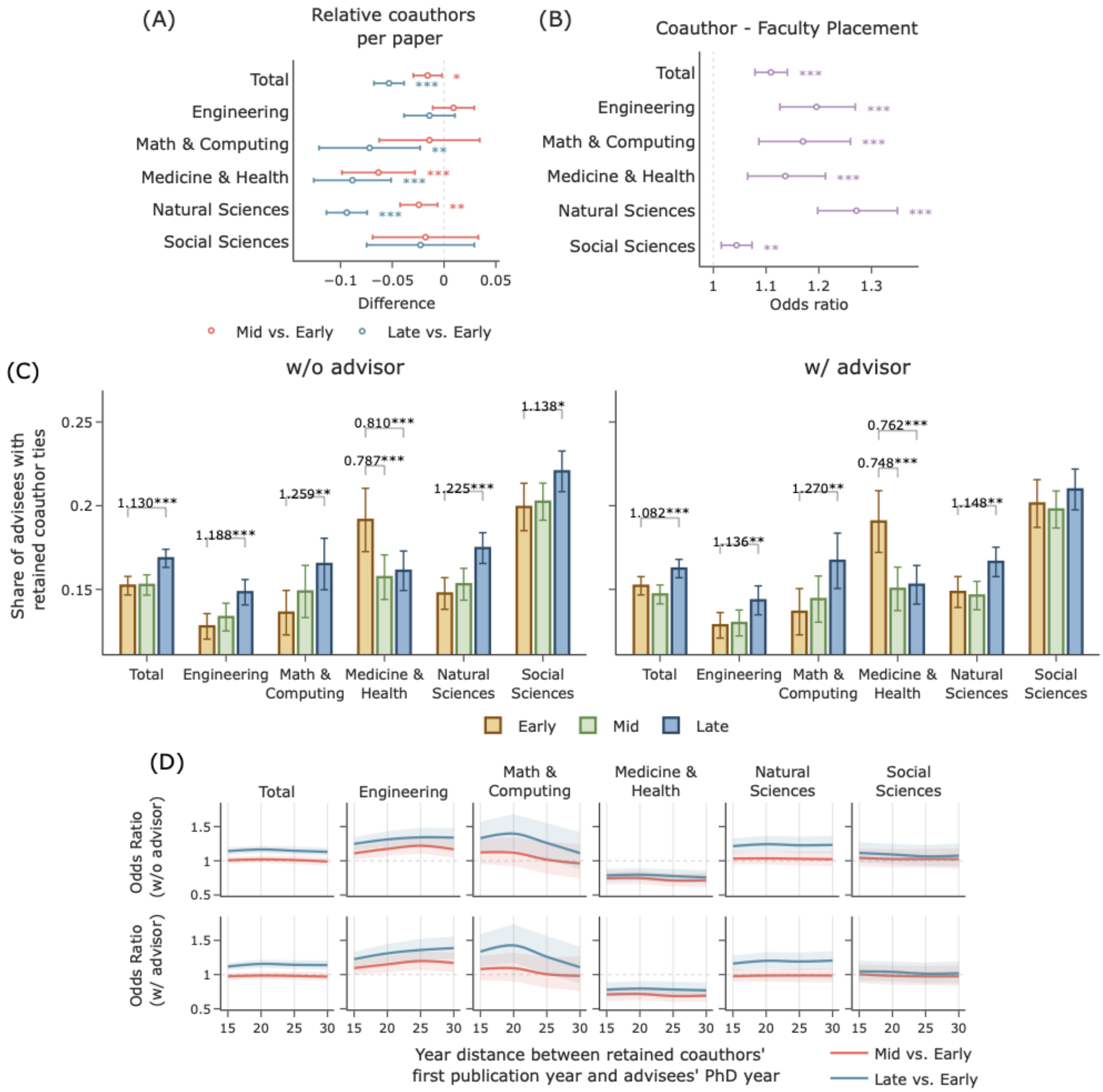

**Supplementary Figure 4-6: Robustness check using untrimmed data.**

**Supplementary Figure 4:** See Figure 1 for caption.

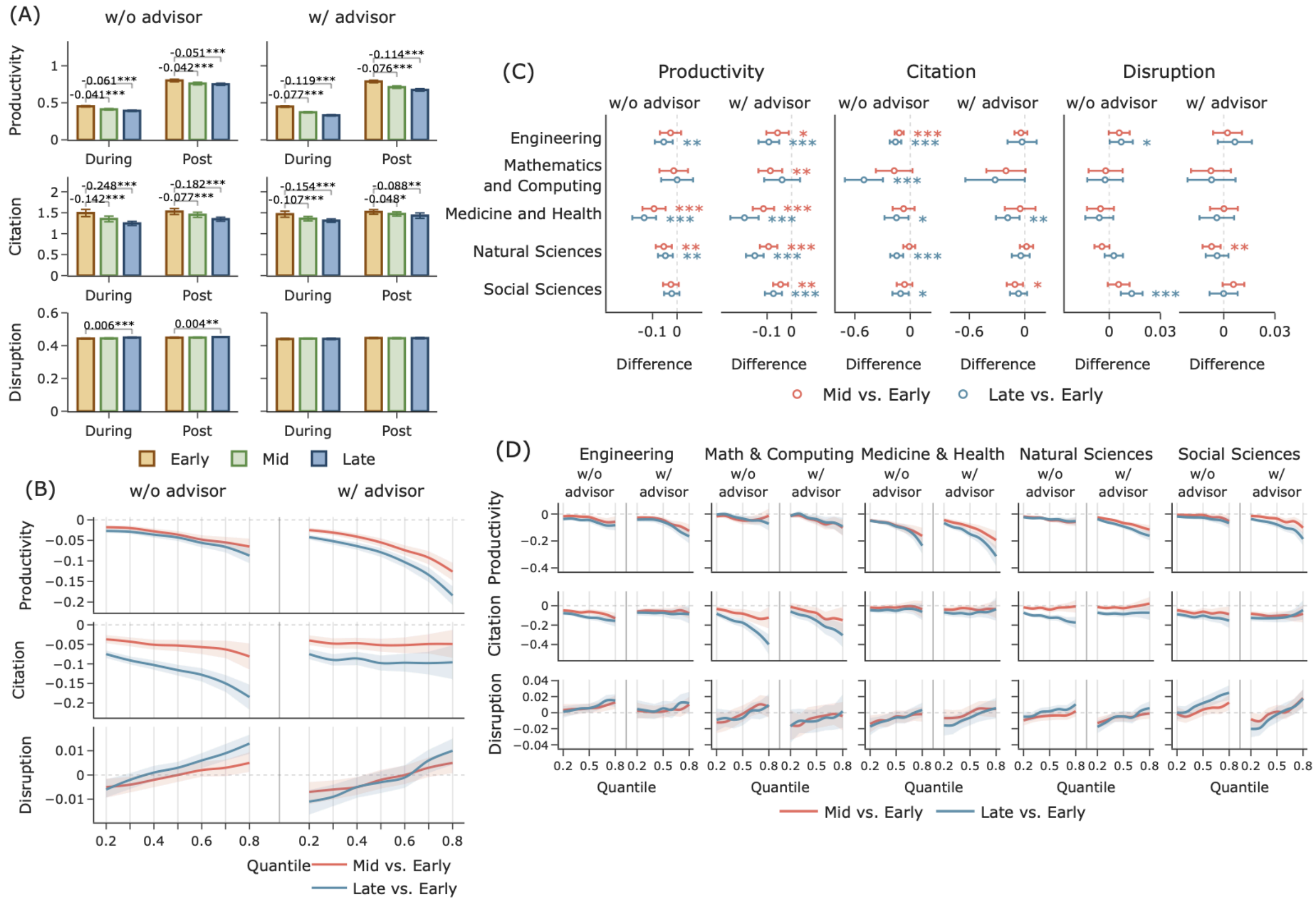

**Supplementary Figure 5:** See Figure 2 for caption.

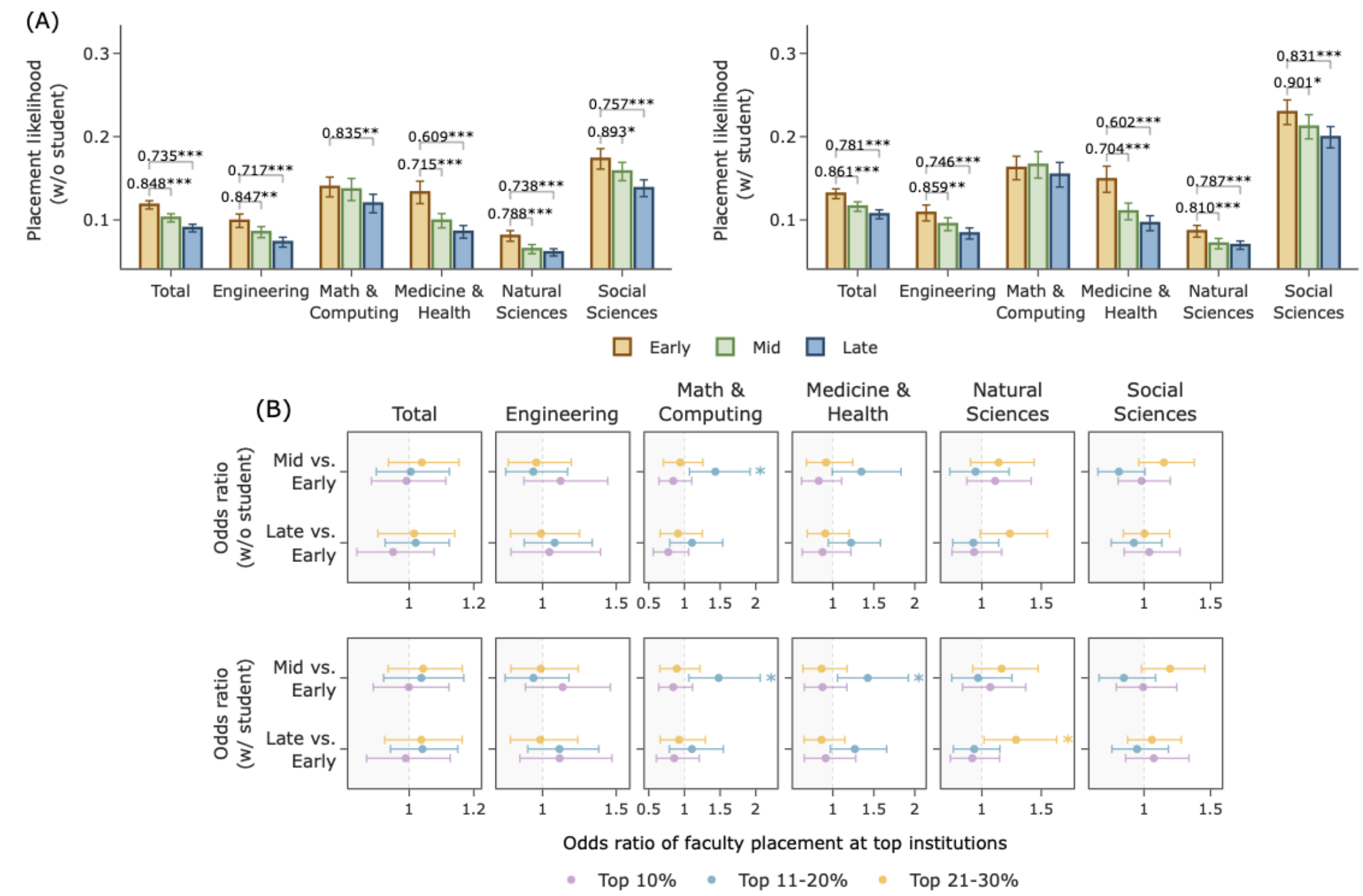

**Supplementary Figure 6:** See Figure 3 for caption.

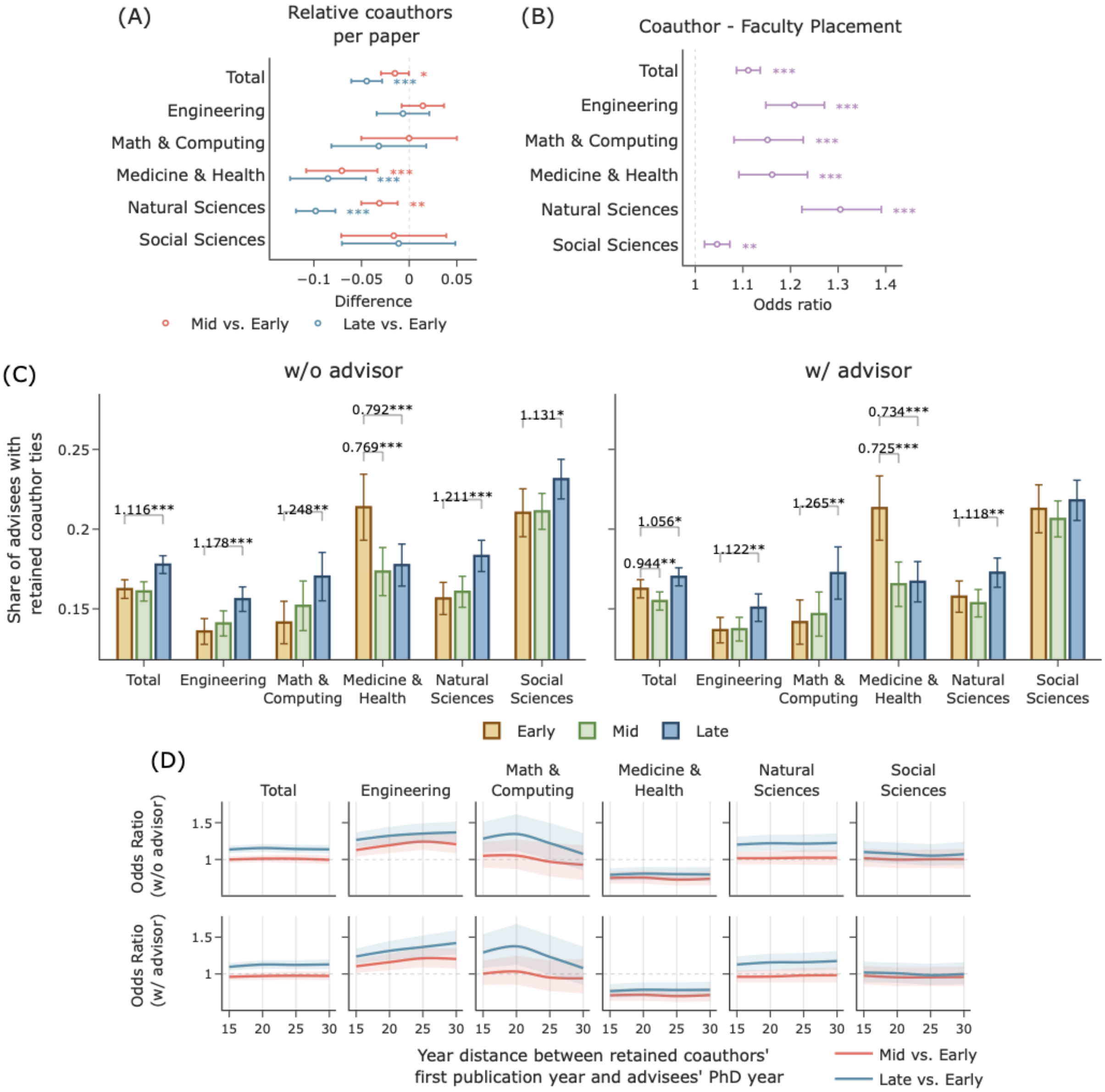

**Supplementary Figure 7-9: Robustness check for the three-class advisor career stage classification using the pre-tenure and post-tenure classification.**

**Supplementary Figure 7:** See Figure 1 for caption.

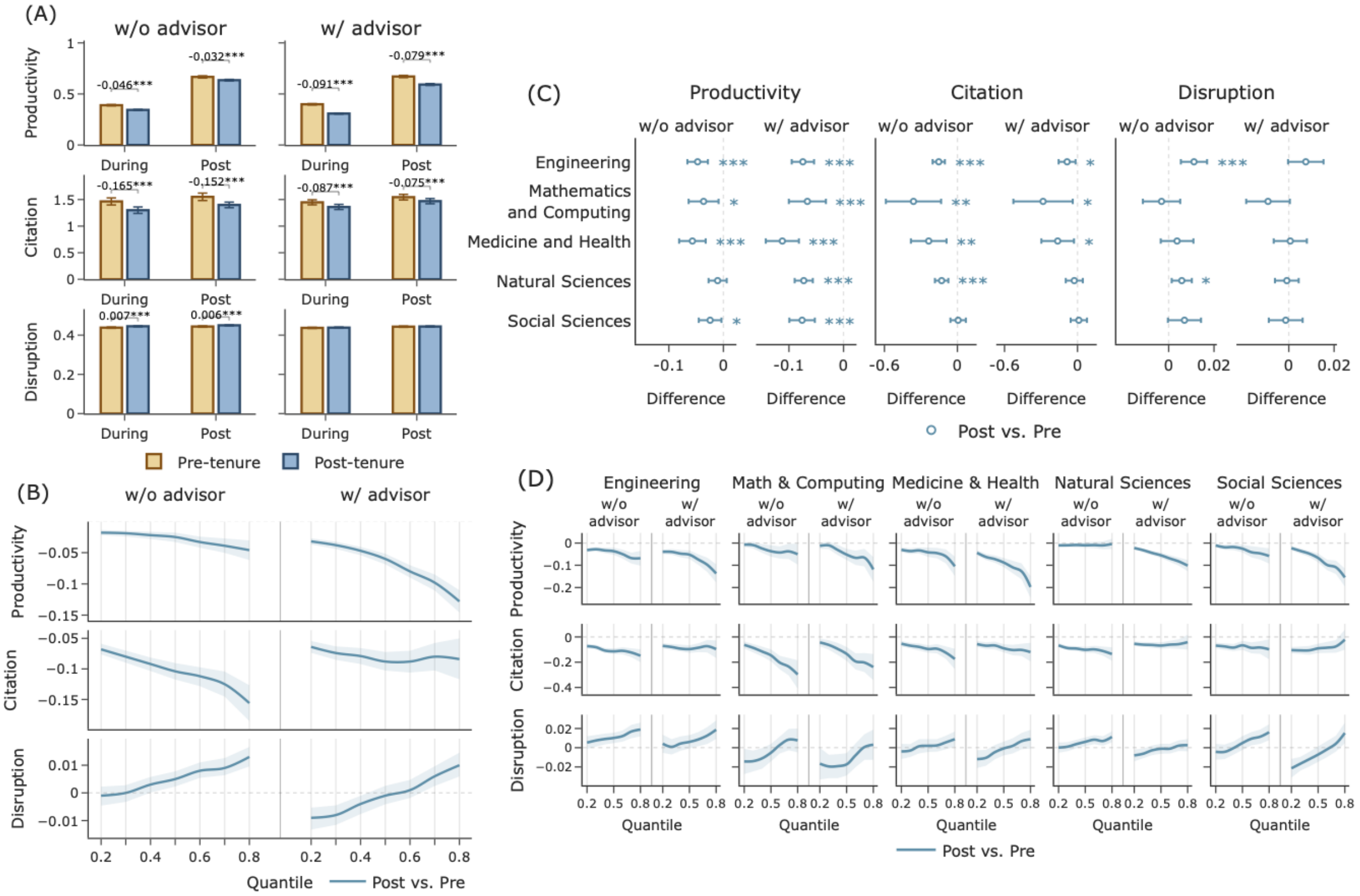

**Supplementary Figure 8:** See Figure 2 for caption.

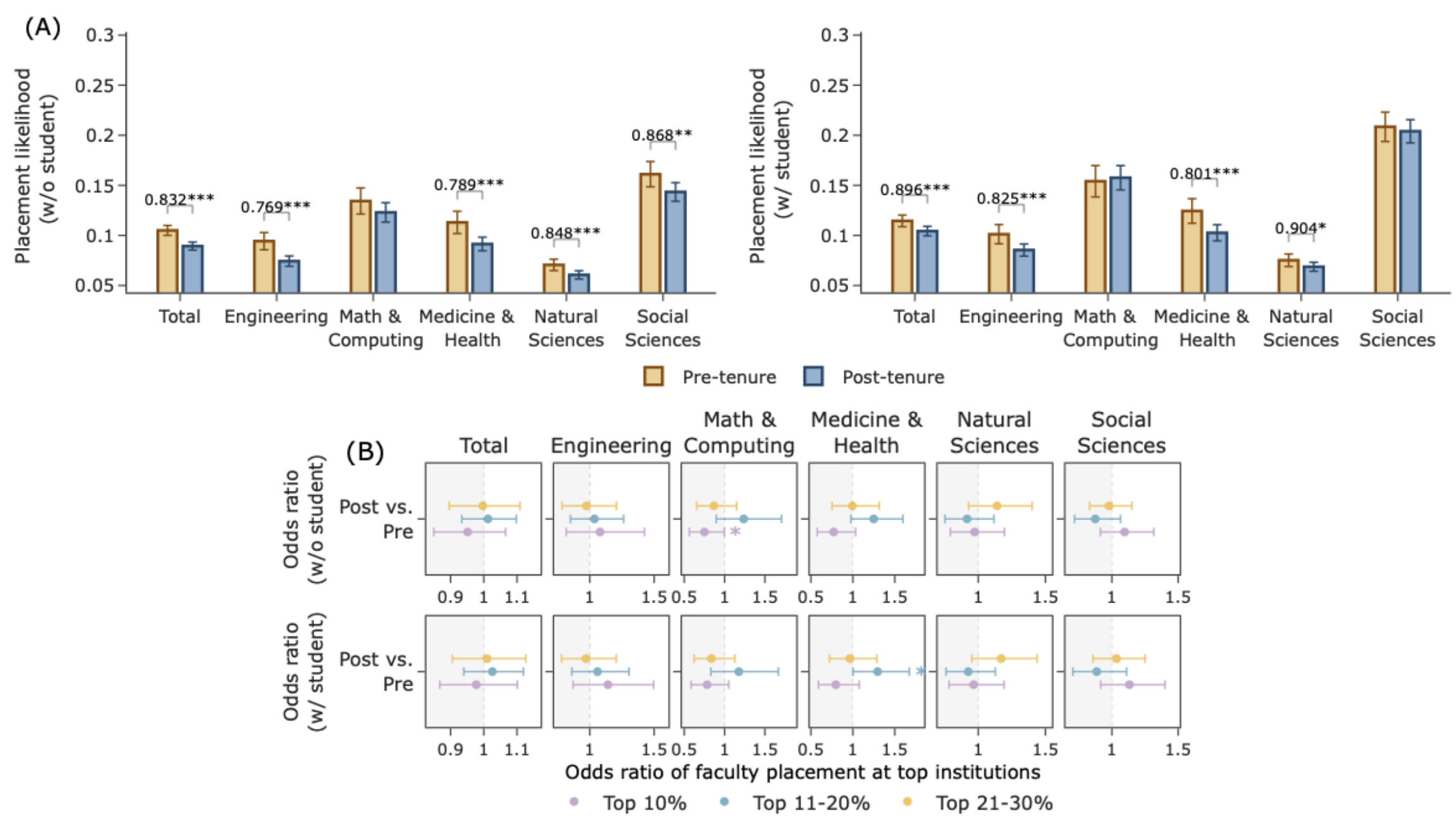

**Supplementary Figure 9:** See Figure 3 for caption.

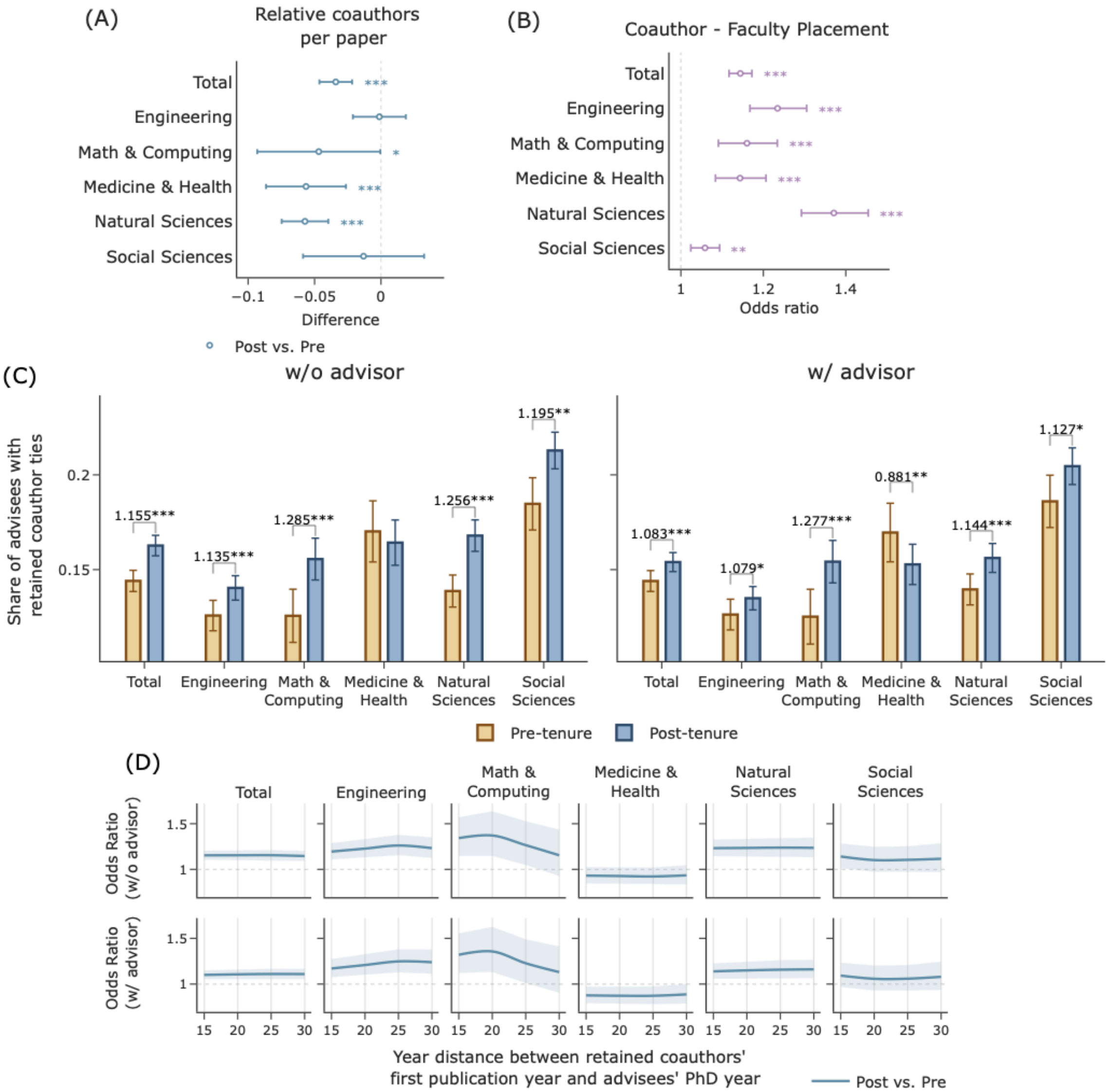

**Supplementary Figure 10:** Differences in the raw number of hit papers between advisees mentored by mid-/late-career advisors and those mentored by early-career advisors. See Figure 1 for caption.

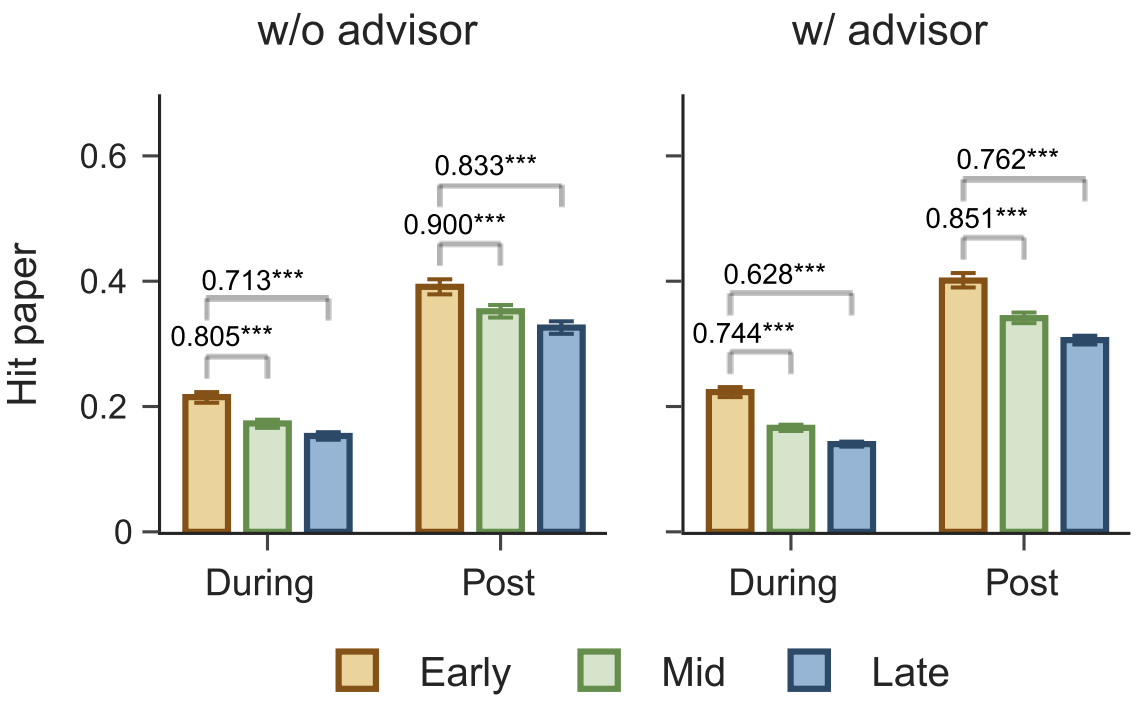

**Supplementary Figure 11:** Differences in the collaborator ties retention proportion between advisees mentored by mid-/late-career advisors and those mentored by early-career advisors. See Figure 3 for caption.

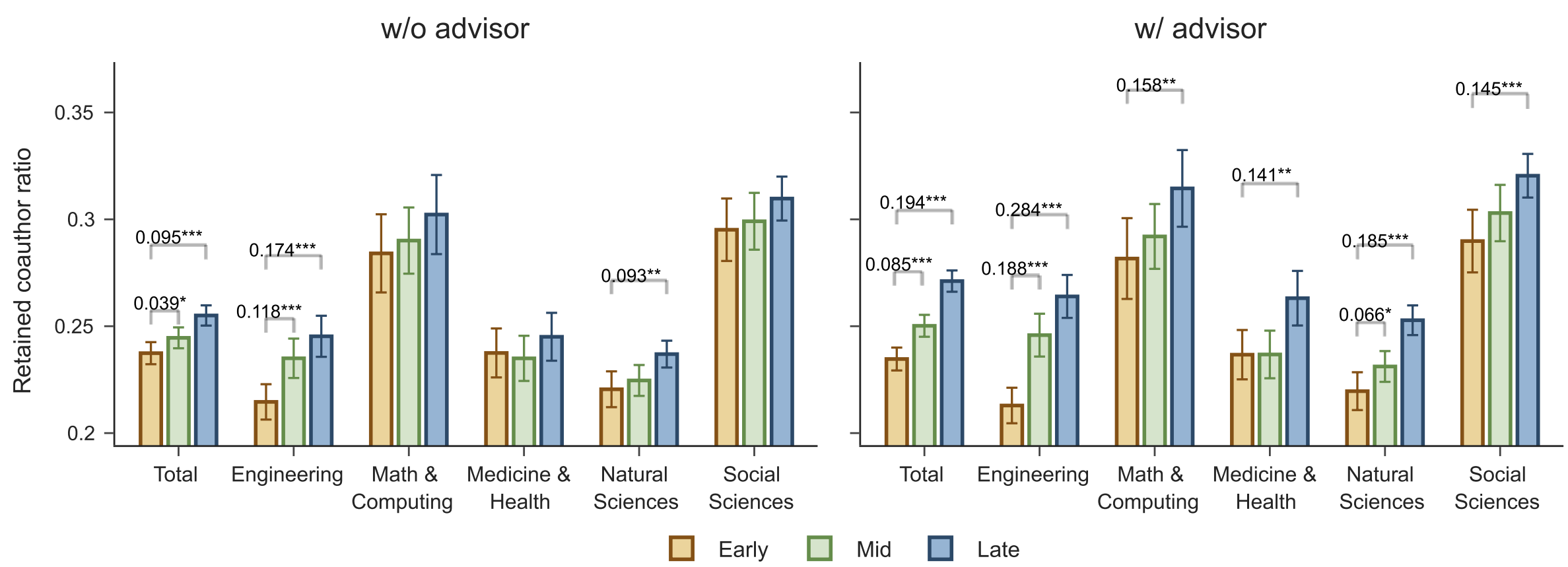